\documentclass[aps,prd,twocolumn,showpacs,superscriptaddress,preprintnumbers, floatfix,nofootinbib]{revtex4-1}
\usepackage{amsmath}
\usepackage{hyperref, url}
\usepackage{graphicx,subfigure}
\usepackage[normalem]{ulem}
\usepackage{epsfig}
\usepackage{color}
\usepackage{multirow}
\usepackage{array}
\usepackage[utf8]{inputenc}
\usepackage[T1]{fontenc}
\usepackage[dvipsnames]{xcolor}
\usepackage{comment}
\usepackage{aas_macros}
\hypersetup{colorlinks   = true,
            urlcolor     = blue,
            citecolor    = blue,
            linkcolor    = blue,
            menucolor    = blue,
            anchorcolor  = blue,
            filecolor    = blue}

\enlargethispage{\baselineskip}

\newcommand{\MScomment}[1]{}
\renewcommand{\MScomment}[1]{{\textcolor{red}{\bf [MS: #1]}}}

\usepackage{graphicx}

\newcommand{\nn}{\nonumber}

\def\a{\alpha} \def\b{\beta}  \def\d{\delta}  
    
 \def\k{\kappa}  \def\m{\mu} \def\n{\nu}

 \def\D{\Delta}  \def\L{\Lambda}

 \newcommand{\Lcal}{{\mathcal L}}

\begin{document}
\preprint{\hfill FERMILAB-PUB-21-687-T-V}
\preprint{\hfill IPMU21-0085}

\title{SpaceQ -- Direct Detection of
Ultralight Dark Matter with Space Quantum Sensors}

\author{Yu-Dai Tsai}
\email{yt444@cornell.edu}
\affiliation{Department of Physics and Astronomy,
University of California, Irvine, CA 92697-4575, USA}
\affiliation{Fermi National Accelerator Laboratory (Fermilab), Batavia, IL 60510, USA}
\affiliation{Kavli Institute for Cosmological Physics (KICP), University of Chicago, Chicago, IL 60637, USA}

\author{Joshua Eby}
\email{joshaeby@gmail.com}
\affiliation{Kavli Institute for the Physics and Mathematics of the Universe (WPI), \mbox{The University of Tokyo Institutes for Advanced Study, The University of Tokyo, Kashiwa, Chiba 277-8583, Japan}}

\author{Marianna S. Safronova}
\email{msafrono@udel.edu}
\affiliation{Department of Physics and Astronomy, University of Delaware, Newark, Delaware 19716, USA}
\affiliation{Joint Quantum Institute, National Institute of Standards and Technology and the University of Maryland, College Park, Maryland 20742, USA}

\date{\today}

\begin{abstract}
\noindent 

Recent advances in quantum sensors, including atomic clocks, enable searches for a broad range of dark matter candidates. The question of the dark matter distribution in the Solar system critically affects the reach of dark matter direct detection experiments. Partly motivated by the NASA Deep Space Atomic Clock (DSAC), we show that space quantum sensors present new opportunities for ultralight dark matter searches, especially for dark matter states bound to the Sun. We show that space quantum sensors can probe unexplored parameter space of ultralight dark matter, covering theoretical relaxion targets motivated by naturalness and Higgs mixing. If an atomic clock were able to make measurements on the interior of the solar system, it could probe this highly sensitive region directly and set very strong constraints on the existence of such a bound-state halo in our solar system. We present sensitivity projections for space-based probes of ultralight dark matter which couples to electron, photon, and gluon fields, based on current and future atomic, molecular, and nuclear clocks.

\end{abstract}

\maketitle

\section{Introduction}

In addition to explaining the dark matter (DM) of the universe, ultralight dark matter (ULDM) can be motivated by naturalness  \cite{Graham:2015cka,Banerjee:2018xmn}, string theory~\cite{Svrcek:2006yi, Arvanitaki:2009fg, Cicoli:2012sz, Visinelli:2018utg}, and dark energy~\cite{Peccei:1987mm,Wetterich:1987fm,Ratra:1987rm,Wetterich:2002ic,Khoury:2003rn,Kim:2002tq, Ibe:2018ffn, Choi:2021aze}. The ``fuzzy'', wavelike nature of such particles can also affect structure formation~\cite{Hu:2000ke, Hui:2016ltb, Mocz:2019pyf}. 
An important probe of ultralight dark matter arises in precision tests using atomic clocks and other quantum technologies \cite{PhysRevD.50.3614,Hoyle:2004cw,Williams:2004qba,Mota:2006fz,Brax:2007vm,Schlamminger:2007ht,Brax:2011hb,Wagner:2012ui,Burrage:2014oza,Foot:2014osa,Roberts:2017hla}, 
which are complementary to interesting astrophysical \cite{Jain:2012tn,Arvanitaki:2014wva,Giannotti:2015kwo,Foot:2016wvj,Caputo:2017zqh,Baryakhtar:2017ngi,Croon:2020oga}, cosmological \cite{Hlozek:2014lca,Baumann:2015rya,DEramo:2018vss,Ade:2018sbj,Vagnozzi:2019ezj}, planetary and space  probes \cite{Pitjev:2013sfa,Tsai:2021irw,Poddar:2020exe,Poddar:2021ose}.

Space quantum technologies are known to have important practical applications, including the auto-navigation of spacecrafts, relativistic geodesy \cite{Puetzfeld:2019kki}, linking Earth optical clocks \cite{Gozzard:2021wkf},  secure quantum communications \cite{Yin17}, and others.
The NASA Deep Space Atomic Clock (DSAC) mission has recently demonstrated a factor of $10$ improvement over previous space-based clocks \cite{Burt2021}, and similar or better sensitivity has been achieved by the other atomic clocks in space \cite{CAC2018}.
We aim to demonstrate a new window of opportunity to study ultralight dark matter with such technologies, taking advantage of these and upcoming space missions to study DM in environments that are drastically different from that of the Earth.

In this paper, we study an exciting new avenue of probing ultralight dark matter with future high-precision atomic, molecular, and nuclear clocks\footnote{We sometimes refer to them as \emph{quantum clocks} for simplicity.} in space.
The oscillations of the ultralight dark matter field can induce a time-varying contribution to fundamental constants, including the electron mass and fine-structure constant \cite{Arvanitaki:2014faa,VanTilburg:2015oza}. Exceptional enhancements of DM density that can be enabled by the bound halos present an opportunity for direct DM detection with clocks \cite{Banerjee:2019epw,Banerjee:2019xuy}.

We propose a clock-comparison satellite mission with two clocks onboard, to the inner reaches of the solar system to search for the dark matter halo bound to the Sun, probe natural relaxion parameter space, and look for the spatial variation of the fundamental constants associated with a change in the gravitation potential. We show that the projected sensitivity of space-based clocks for detection of Sun-bound dark matter halo exceeds the reach of Earth-based clocks by orders of magnitude. We consider both the projected bounds for the clocks that were already demonstrated, and the novel nuclear and molecular clocks under development. This proposal of a clock-comparison experiment in a variable-gravity environment can test the potential spatial variations of fundamental constants under the change in the gravitational potential \cite{Safronova:2017xyt,Lange:2020cul}. We show that using space-based quantum clocks, one can improve the precision by two orders of magnitudes for this measurement in comparison to similar tests on Earth or near-Earth orbits. We also discuss other new physics searches enabled by clock-comparison experiments in space.

Unless otherwise specified, we use the convention of natural units ($\hbar = c = 1$) in this work.

\section{Quantum Clock Searches for ULDM}

ULDM scalar field couplings to the Standard Model can induce oscillations of fundamental constants, including masses and couplings.
Consider the following interaction Lagrangian for a DM scalar field $\phi$:
\begin{equation} \label{eq:Lag}
 \Lcal \supset \k \phi \left(d_{m_e}m_e\bar{e} e 
 + \frac{d_\a}{4}F_{\m\n}F^{\m\n}
 + \frac{d_g\,\b_3}{2 g_s}G^{A}_{\m\n}G^{A\m\n}\right),
\end{equation}
where $e$ is the electron field, $F^{\m\n}$ ($G^{A\m\n}$) is the electromagnetic (QCD) field strength, $g_s$ and $\beta_3$ are the strong interaction coupling constant and beta function (respectively), and $\k = \sqrt{4\pi}/M_P$ with $M_P=1.2\times10^{19}$ GeV. In a DM background field of amplitude  $\phi=\phi_0$, the couplings in Eq.~\eqref{eq:Lag} induce modification of the electron mass $m_e$, fine-structure constant $\alpha$, and strong coupling $\alpha_s \equiv g_s^2/4\pi$, respectively. However, the fundamental oscillatory nature of the ULDM field implies that the contribution to these parameters is oscillatory as well, oscillating at the DM Compton
frequency $\omega = m_\phi c^2/\hbar$. 

Atomic physics experiments, including atomic clock-comparison tests, have shown great promise to probe these signals. The very low fractional uncertainty in frequency that has been achieved corresponds to similar sensitivity to the oscillatory signals of the form
\begin{align} \label{eq:mualpha}
 \mu(\phi) \simeq \mu_0\left(1 + d_{m_e}\k\phi\right), \quad
 \alpha(\phi) \simeq \a_0 \left(1 - d_\a \k\phi\right) \nn \\
 \alpha_s(\phi) \simeq \alpha_{s,0}\left(1 - \frac{2d_g \beta_3}{g_s}\k\phi\right),
\end{align}
where $\mu=m_e/m_p$ is the electron-proton mass ratio, and the subscript ``0'' denotes the central (time-independent) value of $\mu$, $\a$, and $\a_s$. Variation of the strong coupling $\alpha_s$ gives rise to variation of the dimensionless ratio \cite{Damour:2010rm,Damour:2010rp}
\begin{equation}
 \left(\frac{m_q}{\Lambda_{\rm QCD}}\right)(\phi) \simeq 
 \left(\frac{m_q}{\Lambda_{\rm QCD}}\right)_0\left(1-d_g\k\phi\right).
\end{equation}
where $\L_{\rm QCD}$ is the QCD scale and $m_q$ is the averaged light quark mass.
There are now many dedicated experiments searching for these types of signals \cite{VanTilburg:2015oza,Hees:2016gop,Wcislo2018,Aharony:2019iad,Antypas:2019qji,Kennedy:2020bac,Savalle:2020vgz,Oswald:2021vtc}.

Atomic clock accuracy has improved immensely over the past decade, and so too has their ability to test variation of fundamental constants; we review recent work in this field in Section \ref{sec:clocks}.
Other probes of ultralight scalar fields include equivalence principle (EP) tests, which do not need to assume anything about the DM density in the vicinity of the experiment, as they search for virtual exchange of $\phi$ particles that appear as a ``fifth force" not proportional to $1/r^2$ (see e.g. \cite{Wagner:2012ui,Berge:2017ovy,Hees:2018fpg}).
Historically, EP tests have outperformed atomic physics probes across a wide range of ULDM mass parameters, with the exception of very light particles $m_\phi \lesssim10^{-17}$ eV \cite{Kennedy:2020bac}, at least for generic couplings and under the usual assumption of $\rho_{DM} = 0.4$ GeV/cm$^3$ for the local density of dark matter.
On the other hand, atomic physics probes couple directly to the DM density and, therefore, allow for direct detection. Furthermore, such experiments have the ability to probe bound-state DM in our solar system, as we will explain below, and a space-based clock allows one to probe novel parameter space as well.
Future development of the nuclear clock, expected to be  $10^4-10^5$ times more sensitive to variations of $\alpha$ than all operating atomic clocks, will drastically increase the discovery reach of such experiments. In addition, the nuclear clock has strong sensitivity to $m_q/\Lambda_{\rm QCD}$; for further details, see Section \ref{sec:clocks}.

\section{Solar System Halos}

The local DM density $\rho_{\rm DM}$ is a key parameter dictating experimental sensitivity; on the basis of halo modeling and (weak) local constraints (see below, as well as Appendix \ref{app:constraints}), its value is typically assumed to be $\rho_{DM} = 0.4$ GeV/cm$^3$. For ultralight particles, the field oscillates coherently on a timescale dictated by the virial velocity $v_{\rm DM} \simeq 10^{-3}c$, given by $\tau_{\rm coh} \approx 2\pi (m_\phi\,v_{\rm DM}^2/\hbar c^2)^{-1} \simeq 2\pi\times 10^6\hbar/m_\phi$. 
However, the possibility that a large density of such fields could become bound to objects in the solar system has been considered, which would give rise to unique signals and strongly modified values for the local DM density and timescale of coherent DM oscillations \cite{Banerjee:2019epw,Banerjee:2019xuy,Anderson:2020rdk}. Here we focus on the specific case of a bound ULDM halo around the Sun, known as a \emph{solar halo} (SH).\footnote{Note that a SH shares similarities to (though distinct from) a stellar DM basin \cite{VanTilburg:2020jvl,Lasenby:2020goo}, which is generally formed from heavier particles.}

There are intriguing hints that some density of ULDM would become bound to the Sun. One piece of suggestive evidence arises in numerical simulations of galaxy formation in ULDM with $m_\phi\sim10^{-22}$ eV, which have recently included the presence of (fixed) baryonic gravitational potential \cite{Veltmaat:2019hou}. The simulation suggests that the same relaxation processes that form boson stars in DM-only case can instead form a halo-like configuration, akin to a gravitational atom (analogous to a hydrogen atom ground state with a gravitational potential), in the presence of a baryonic potential. If this holds also at larger $m_\phi$, as we will consider below, it implies a plausible formation mechanism for ULDM to become bound to the Sun.  It has also been suggested that a SH could form from adiabatic contraction during star formation \cite{Anderson:2020rdk}. In this work, we analyze the consequences of the existence of a SH on atomic clock searches for ULDM, with a focus on possibilities for future missions in space; previous work has focused instead on terrestrial searches (e.g. \cite{Banerjee:2019epw}).

A SH can be thought of as similar to a boson star \cite{Kaup:1968zz,Ruffini:1969qy,Colpi:1986ye,Chavanis:2011zi}, but supported by the external gravitational effect of the host body (the Sun) rather than self-gravity. The radius of a SH takes the form \cite{Banerjee:2019epw}
\begin{equation} \label{eq:Rstar}
 R_\star \simeq \frac{M_P^2}{M_{\rm ext}\,m_\phi^2},
\end{equation}
where $M_{\rm ext}=M_\odot$ is the mass of the external host body; note that $R_\star$ is independent of the total mass in the halo $M_\star$.
For ULDM masses of $m_\phi \sim$ few$\times10^{-14}$ eV, the radius of a SH is roughly $1$ AU (the average orbital radius of the Earth), and $R_\star$ grows as $\propto m_\phi^{-2}$ as $m_\phi$ mass decreases. Therefore, terrestrial atomic probes are sensitive only to a lower mass range fixed by the requirement $R_\star \gtrsim $ 1\,AU. 

Note that a bound halo around the Earth would modify signals in the higher mass range $10^{-12}$ eV$\lesssim m_\phi \lesssim 10^{-7}$ eV; we discuss the resulting effects on atomic clocks in orbit around the Earth in Appendix \ref{app:earthclock}.

Space-based atomic clocks are notably different from terrestrial ones regarding sensitivity to ULDM probes. Firstly, a space clock would provide a novel method to probe a SH at larger $m_\phi$, when the radius of the SH in Eq.~\eqref{eq:Rstar} is smaller than $1$ AU. Secondly, and perhaps more strikingly, the constraints on an SH with a small radius are very weak; if an atomic clock were able to make measurements on the interior region of the solar system, it could probe this highly sensitive range directly and set very strong constraints on the existence of such a halo in our solar system.
 Current constraints on a SH in our solar system arise from measurements of solar system ephemerides, especially from Mercury, Mars, and Saturn, which are known with very high precision \cite{Pitjev:2013sfa}. The resulting maximum mass of a SH, following \cite{Banerjee:2019epw}, is of order $10^{-12}M_\odot$ at $m_\phi \simeq 10^{-14}$ eV, and weaker elsewhere; in what follows, we use the full range of gravitational constraints, and we always enforce that $M_\star < M_\odot/2$ as a naive requirement on the total mass in our solar system.  See Appendix \ref{app:constraints} for further details.

We also note that the effective coherence time of the oscillations of the bound ULDM field $\tau_\star$ is generically larger than that of the virialized DM scenario \cite{Banerjee:2019epw,Banerjee:2019xuy}, where $\tau_{\rm DM} \simeq 10^6/m_\phi$. In a narrow range around $m_\phi \simeq 10^{-13}$ eV, however, $\tau_\star < \tau_{\rm DM}$ by a factor of few, possibly reducing the sensitivity reach of atomic clock probes by as much as an order of magnitude; we discuss this further in Appendix \ref{app:constraints}.

\section{Atomic, Molecular, and Nuclear Clocks} \label{sec:clocks}

To detect ultralight dark matter with high-precision clocks, one measures a frequency ratio of two clocks with different sensitivities to the variation of fundamental constants over a period of time \cite{Arvanitaki:2014faa}. The discrete Fourier transform of the resulting time series then allows the extraction of a peak at the dark matter Compton frequency, with an asymmetric lineshape
\cite{Arvanitaki:2014faa,Der18,CenBlaCon20}. The lack of such a signal allows one to establish bounds on the DM parameter space. It is also possible to carry out such a measurement with a single clock by comparing the frequency of atoms to the frequency of the local oscillator (i.e., cavity)
\cite{Wcislo2016,Kennedy:2020bac}.

The present proposal calls for a two-clock or clock-cavity setup onboard a satellite. It does not require a comparison to Earth-bound clocks. There are several factors one has to consider while selecting clocks for a proposed mission: (1) variation of which fundamental constants do we want to probe and what are the corresponding sensitivity factors; (2) what are the clock stabilities and systematic uncertainties; and (3) the difficulty of making these clocks space-ready.

The dimensionless sensitivity factors $K$ of a pair of clocks translate the fractional accuracy of the ratio of frequencies $\nu$ to the fractional accuracy in the variation of the fundamental constant. For example, for the fine-structure constant
\begin{equation}
\frac{\partial}{\partial t} \textrm{ln}\frac{\nu_2}{\nu_1}=(K_2-K_1)\frac{1}{\alpha}\frac{\partial \alpha}{\partial t},
\end{equation}
where indices 1 and 2 refer to clocks 1 and 2, respectively.
If the frequency ratio is measured with relative $10^{-18}$ precision and $\D K \equiv K_2-K_1=1$, then such an experiment will be able to measure the fractional change in $\alpha$ with 10$^{-18}$ precision. If $\D K=10^4$ then the 10$^{-18}$ accuracy of the frequency ratio allows one to detect a change in $\alpha$ at the 10$^{-22}$ level. The sensitivity factors $K$ to $\alpha$-variation for all atomic clocks can be computed from first principles with high precision \cite{FlaDzu09}. They increase for atoms with heavy nuclei and depend on the electronic configurations of the clock states.

At present, all operating atomic clocks are either based on transitions between the hyperfine substates of the ground state of the atom (microwave clocks: H, DSAC Hg$^+$, Rb, Cs) or transitions between different electronic levels (optical clocks: Al$^+$, Ca$^+$, Sr, Sr$^+$, Yb, Yb$^+$, and others) \cite{LudBoyYe15}. The typical frequencies of the optical clock transitions are $0.4-1.1\times10^{15}$~Hz, while the frequencies of the microwave clocks are several orders of magnitude smaller,  a few GHz. The optical clock frequency is only sensitive to the variation of $\alpha$, with varying sensitivity factors $K$. Microwave clocks are sensitive to variation of $\alpha$ and $\mu=m_e/m_p$ ratio (with a sensitivity factor of $K = 1$), and there is also a small sensitivity of microwave clocks to $m_q/\Lambda_{QCD}$.
The sensitivity to the variation of $\alpha$ of most currently operating atomic clocks are small: $K$(Al$^+$)=0.01, $K$(Ca$^+$)=0.1, $K$(Sr)=0.06, $K$(Sr$^+$)=0.4, $K$(Yb)=0.3, $K$(Yb$^+$~E2)=1, with a notable exception of Yb$^+$ clock based on the octuple transition with $K=-6$ \cite{FlaDzu09}.  For the microwave Cs clock, $K=2.83$.
The most recent limits on the slow drifts of $\alpha$ and $\mu$ are given in \cite{LanHunRah21}.
Comparing any optical clock to cavity gives $\Delta K=1+K_{\textrm{clock}}$, where $K_{\textrm{clock}}$ is given above.

There are two characteristics to consider when evaluating the state-of-the-art clocks: stability and uncertainty \cite{LudBoyYe15,PolOatGil13}. \emph{Stability} is the precision with which we can measure a quantity. It is usually determined as a function of averaging time, since noise is reduced through averaging for many noise processes, and the precision increases with repeated measurements. Probing the resonance using the Ramsey method of separated fields\footnote{Ramsey scheme involves applying two $\pi/2$ laser pulses with a wait (free precession) time in-between. The $\pi/2$ laser pulse creates a superposition of two clock states.} in the regime where the stability is limited by fundamental quantum projection noise, a clock instability is limited by \cite{LudBoyYe15,PolOatGil13}
\begin{equation}
\sigma(\tau)\approx \frac{1}{2\pi \nu_0 \sqrt{N T_m {\rm min}(\tau,\tau_\star)}},
\label{sigma}
\end{equation}
where $\nu_0$ is the clock transition frequency, $N$ is the number of atoms or ions used in a single measurement, $T_m \sim 2 \pi \delta \nu$  is the maximum possible time of a single measurement cycle (i.e., the free-precession time where $\delta \nu$ is the spectroscopic linewidth of the clock transition), and $\tau$ is the averaging period \cite{LudBoyYe15,PolOatGil13}. Generally, $T_m$ is still limited by the clock laser coherence rather than the natural linewidth, which represents a fundamental limit. This formula demonstrates the advantages of the optical clocks over the microwave clocks due to five orders of magnitude increase in the clock transition frequency $\nu_0$. Note that per Eq. \eqref{sigma}, the clock stability is diminished when $\tau<\tau_\star$, which becomes relevant at large values of $m_\phi$, e.g. $\tau_\star \lesssim 1$ day for $m_\phi\gtrsim 10^{-13}$ eV; we explain this in greater detail in Appendix \ref{app:constraints}.

The absolute \emph{uncertainty} of an atomic clock describes how well we understand the physical processes that shift the measured frequency from its unperturbed natural value. Several optical clocks have reached uncertainty at the 10$^{-18}$ level \cite{BreCheHan19,SanHunLan19,BotKedOel19}, while microwave clocks are at the 10$^{-16}$ level \cite{WeyGerKaz18}, which is at the achievable technical limit. There is no apparent technical limit to significant further improvement of the optical clocks \cite{BotKenAep21}. Portable high-precision optical clocks were also demonstrated \cite{KelBurKal19,TakUshOhm20}.

We now consider all parameters in Eq.~(\ref{sigma}) in the context of dark matter detection.
All present optical clocks are based on either neutral atoms or singly-charged ions. To achieve high precision, the atoms/ions that serve as the frequency standards have to be trapped, leading to significant differences in atom and ion clock design due to different trapping technologies. Neutral atom clocks are sometimes referred to as ``lattice clocks'', as atoms are held in optical lattices, i.e., light potentials created by counterpropagating laser beams.
While ions trap technology design is simpler than that of lattice clocks, all of the ion clocks are operating with a single ion, i.e., $N = 1$, leading to lower stability compared to the optical lattice clocks that have $N \sim 1000$. Much larger values of $N$ were recently demonstrated for lattice clocks \cite{BotKenAep21}. Development of multi-ion clocks is in progress \cite{KelBurKal19,TanKaeArn19}.
  
The range of dark matter masses for which the DM frequency signal can be extracted without sensitivity loss depends on parameters of the clock operation $T_m$ and $\tau$ and DM coherence time \cite{CenBlaCon20}. Generally, one needs to have at least one DM oscillation during the total measurement time $\tau$, losing the sensitivity beyond this point. For high frequencies, one eventually will have multiple dark matter oscillations during free precession (probe) time $T_m$, leading to a loss of sensitivity. This is particularly significant for this proposal, as the characteristic probe time of $T_m=1$\ sec corresponds to DM mass of $4\times10^{-15}$~eV. This problem can be partially remedied by either reducing $T_m$ (leading to reduced short-term stability) or applying additional ``dynamical decoupling" (DD) series of $\pi$ laser pulses during the clock probe time \cite{Aharony:2019iad}. DD scheme allows to coherently add the DM signal contribution over the probe time and, therefore,  to extract the DM signal that oscillates on a faster timescale than the clock measurement cycle. The DD sequence can be optimized to probe the 10$^{-13}$ eV mass range of specific interest to this work. Here, we assume an ideal case where the clock probes are continuous; realistically, clocks have a ``dead time'' required to prepare the system, which can be  50\% of the entire clock measurement cycle or more. A zero-dead-time clock has been demonstrated with two atomic ensembles \cite{SchBroMcG16}.
  
  \begin{figure*}
\centering
\begin{minipage}[c]{0.45\textwidth}
\hspace{-7cm} \large{(a)} \\
\vspace{-0.5cm}
\includegraphics[width=\linewidth]{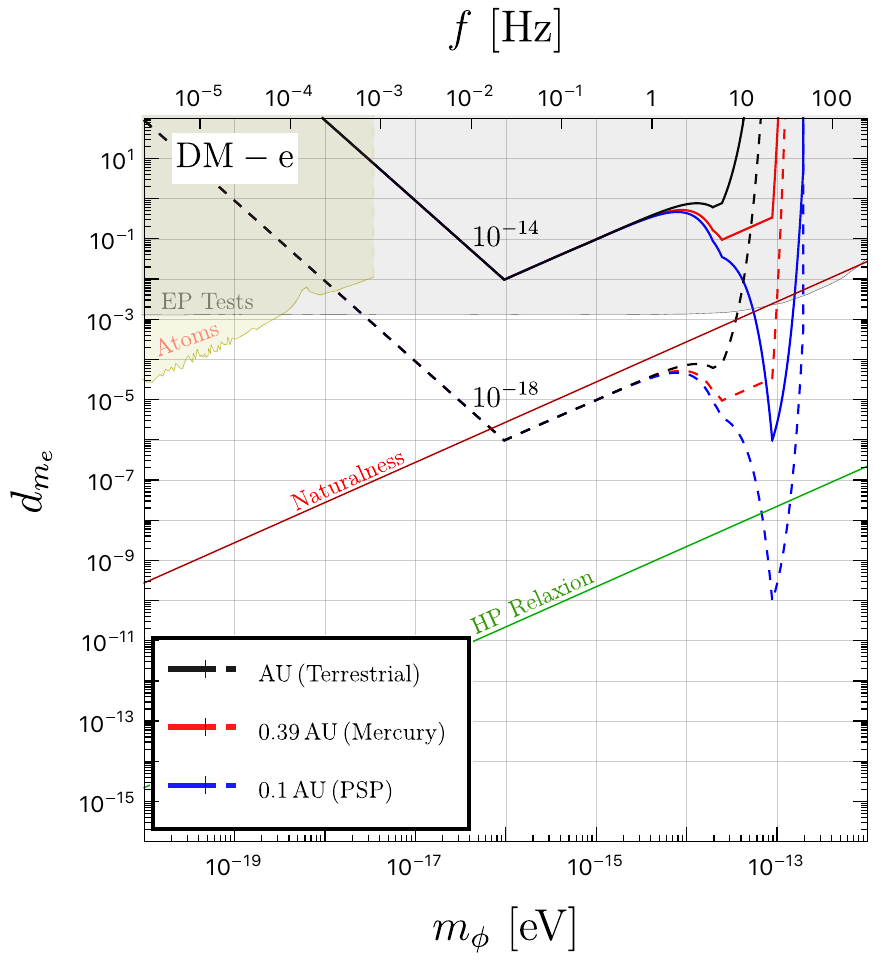}
\end{minipage}
\hfill
\begin{minipage}[c]{0.45\textwidth}
\hspace{-7cm} \large{(b)} \\
\vspace{-0.5cm}
\includegraphics[width=\linewidth]{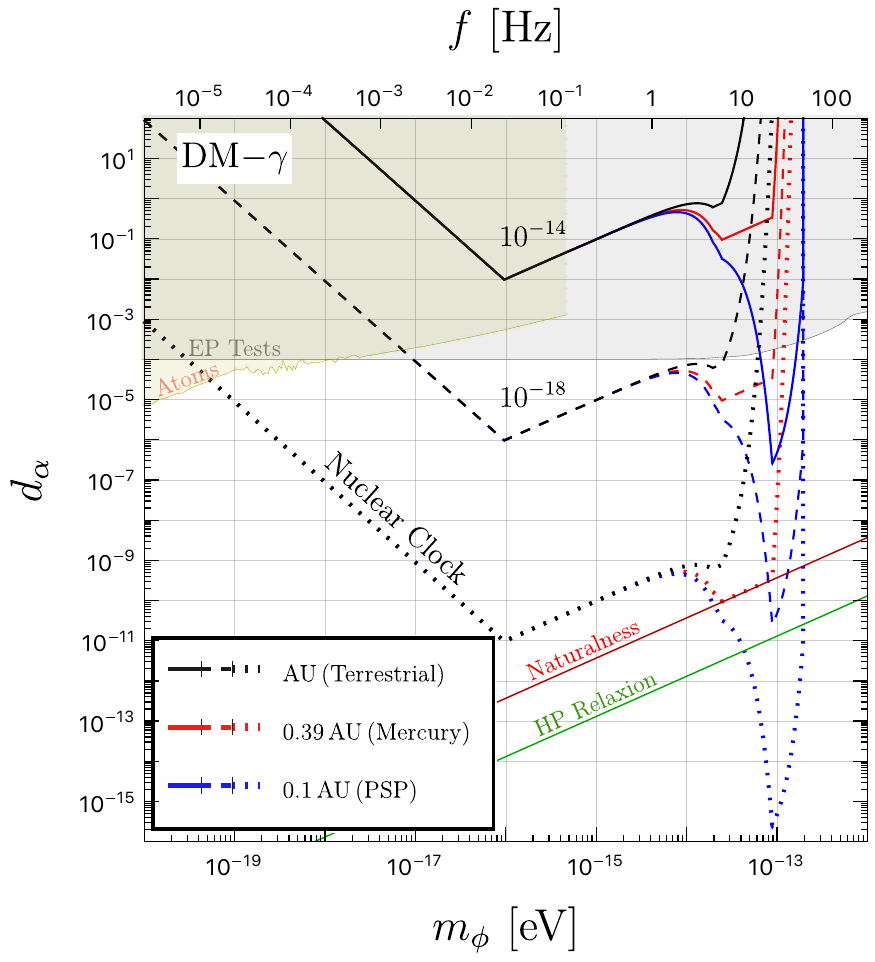}
\end{minipage}
\\
\vspace{0.2cm}
\begin{minipage}[c]{0.45\textwidth}
\hspace{-7cm} \large{(c)} \\
\vspace{-0.5cm}
\includegraphics[width=\linewidth]{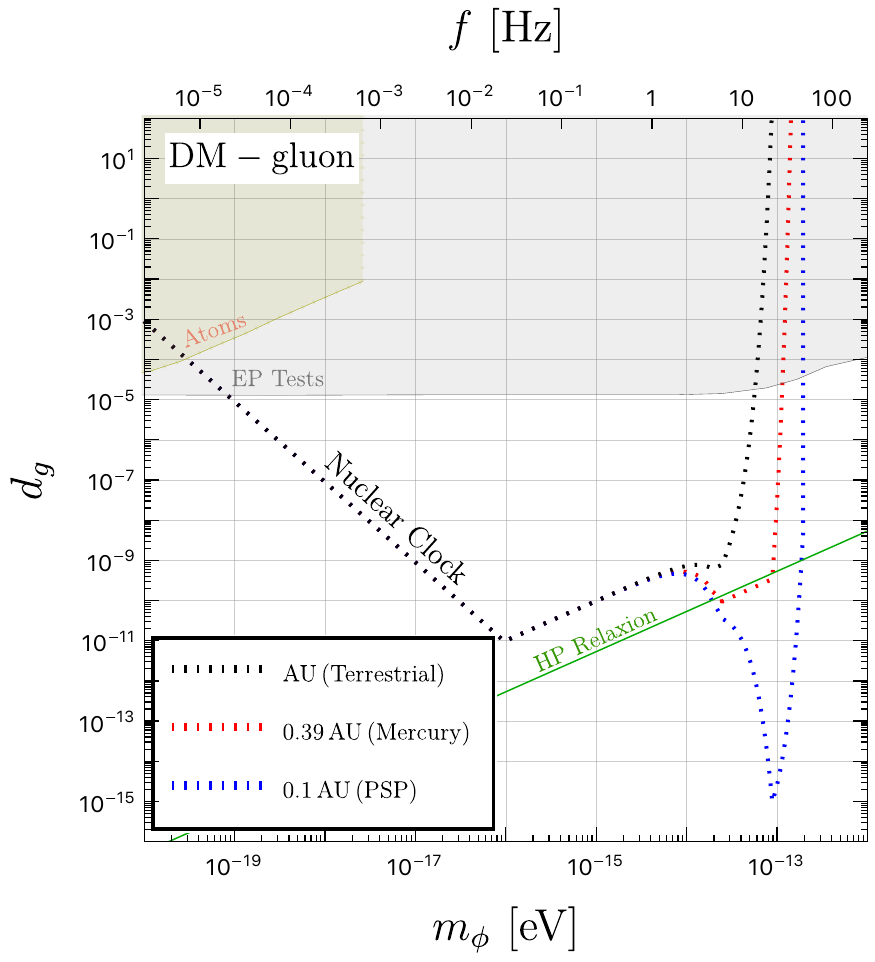}
\end{minipage}
\hfill
\begin{minipage}[t]{.45\textwidth}
\vspace{-4cm}
\caption{\label{fig:sens}
Estimated sensitivity reaches for ultralight dark matter, coupled via Eq. \eqref{eq:Lag} and bound to the Sun. The blue, red, and black denote sensitivity for probes at the distance of 0.1 AU, probes at the orbit of Mercury, and for terrestrial clocks, respectively; note that distances of $r<0.1$ AU have been reached by the NASA Parker Solar Probe mission, reaching $0.06$ AU on its most recent perihelion and aiming for $0.045$ AU at the closest approach \cite{PSP}. Panels (a - c) show projected bounds for the variations of the electron-proton mass ratio $\mu$ (panel a), fine structure constant $\a$ (panel b), and ratio $m_q/\Lambda_{\rm QCD}$ (panel c).
The thick (dashed) lines correspond to assumed experimental sensitivity of $10^{-14}$ ($10^{-18}$) for panels (a) and (b).
The dotted lines in panels (b) and (c) represent the projection for a clock-comparison experiment at the $10^{-19}$ level involving a nuclear clock, $10^4$ sensitivity factor is assumed for a nuclear clock (see Section \ref{sec:clocks} and e.g. \cite{FadBerFla20}).
The gray and yellow shaded regions denote the current constraint from equivalence principle tests \cite{Wagner:2012ui,Berge:2017ovy,Hees:2018fpg} and atomic physics probes of the local DM density $\rho_{DM}$ \cite{VanTilburg:2015oza,Hees:2016gop,Wcislo2018,Kennedy:2020bac} (respectively); the diagonal burgundy and green solid lines denote motivated theory targets \cite{Flacke:2016szy,Choi:2016luu}.
}
\end{minipage}
\end{figure*}

In addition to presently operating clocks, a number of new clocks are being developed, bases on molecules and molecular ions \cite{SafBudDem18,HanKuzLun21}, highly-charged ions (HCIs) \cite{KozSafCre18}, and $^{229}$Th nucleus \cite{PeiSchSaf21}. A lattice clock based on the $4f^{14}6s6p\, ^3P_0 - 4f^{13}6s^25d \,J=2$ transition   in neutral Yb was proposed with $K=15$ ~\cite{SafPorSan18}. Molecular clocks are projected to reach $10^{-18}$ uncertainties \cite{HanKuzLun21}.  Highly-charged ion clocks  and a nuclear clock are estimated to achieve $10^{-19}$ uncertainties and have much higher sensitivities to $\alpha$, $K\sim 100$ for HCIs and $K = -(0.82 \pm 0.25)\times10^4$ \cite{FadBerFla20}  for a nuclear clock, with actual sensitivity to be determined with aid of future measurements of nuclear properties. HCI clocks have to operate in a cryogenic 4K environment, complicating space deployment. Nuclear clocks can be operated as a trapped ion or a solid state clock. Molecular clocks provide sensitivity to $m_e/m_p$ variation and a nuclear clock is highly sensitive to hadronic sector, with possible $K=10^4$ sensitivity to the variation of $m_q/\Lambda_{QCD}$ \cite{Fla06,FadBerFla20}.

In summary, a wide variety of clocks can be selected for a pair of co-located mission clocks. A very attractive possibility is to use Yb$^+$ that has two clock transitions in the same ion giving $\Delta K=7$ \cite{LanHunRah21} (the probe sequence will alternate between two transitions). Such a scheme removes uncertainty due to gravitational potential and temperature differences between clock locations. Development of a two-transition Yb lattice clock proposed in \cite{SafPorSan18} would have the same benefits and provide high stability enabled by thousands of trapped neutral atoms combined with high sensitivity $\Delta K=15$. Comparing  Sr clock \cite{BotKenAep21} to a ultra-stable cavity \cite{Kennedy:2020bac} ($\Delta K=1$) would utilize extraordinary clock stability  but increasing cavity performance will require a cryogenic setup. Future development of a high-precision nuclear clock will enable an ultimate experiment with the highest potential discovery reach.

\section{Sensitivity Reach}

We estimate the sensitivity of a space-based quantum clock to the oscillation of fundamental constants, originating in ULDM fields of mass $m_\phi$ bound to the Sun.
In Fig. \ref{fig:sens}, we estimate the reach for oscillations of $\mu$, $\alpha$, and $m_q/\L_{\rm QCD}$ (through the couplings in Eq.~\eqref{eq:Lag}) in panels (a), (b), and (c), respectively.
As input, we take the possible distances from the Sun of $r=$ 1\,AU (terrestrial searches), $r=0.39$ AU (the orbital radius of Mercury), and the far-future potential for a probe at $r=0.1$ AU; this latter distance is used as a demonstration, and we note that NASA Parker Solar Probe (PSP) has already reached this inner orbit and in fact even nearer to the Sun, reaching $0.06$ AU on its most recent perihelion and aiming for $0.045$ AU at the closest approach \cite{PSP}.
The NASA Parker Solar Probe instruments are designed to study particles and electromagnetic fields for its scientific missions.
Operating atomic clocks in an extreme environment within the Mercury orbits is a subject for future investigation. We note that the Mercury Laser Altimeter Instrument demonstrated a successful laser operation for the MESSENGER mission \cite{2007SSRv..131..451C}.

We observe that for probes in this inner region of the solar system, there is a clear ``peak'' in the sensitivity reach around $m_\phi\simeq 10^{-13}$ eV, which roughly corresponds to the point where $R_\star \simeq 0.1$ AU; at larger $m_\phi$, the exponential cutoff of the SH density function rapidly diminishes the sensitivity (see Eq. \eqref{eq:rhostar}).

In Fig. \ref{fig:sens} (a) and (b), we assume a space-based clock-comparison with accuracy at the level of $10^{-14}$ (thick lines) or $10^{-18}$ (dashed lines);
the former represents only a factor $\sim$few improvement compared to what has already been demonstrated in DSAC \cite{DSAC2021}, and the latter is already achievable  for the variation of $\alpha$ in terrestrial optical clock-comparison experiments (see Section \ref{sec:clocks}, and e.g. \cite{BreCheHan19,SanHunLan19,BotKedOel19}). Future molecular and molecular ions clocks are projected to reach $10^{-18}$ sensitivity to the variation of $\mu$  \cite{HanKuzLun21}.  We observe that even a sensitivity of $10^{-14}$ to these oscillations allows one to probe interesting model space in a narrow range around $m_\phi \sim 10^{-13}$ eV, and space clocks at the $10^{-18}$ level could exceed EP probes over a wide range $3\times10^{-17}$ eV$\lesssim m_\phi \lesssim 2\times10^{-13}$ eV, for a bound SH.

In Fig. \ref{fig:sens} (b) and (c), we include a projection for a clock-comparison experiment at the $10^{-19}$ level involving a nuclear clock (dotted lines), assuming $\D K \simeq 10^4$, which is in line with future projections outlined in Section \ref{sec:clocks} \cite{FadBerFla20}. 

In these estimations, we fix $M_\star$ by the maximal bound-state mass allowed by current constraints, though our projection can be easily rescaled to less optimistic input values using $d^{\rm limit} \propto \phi^{-1} \propto \rho^{-1/2} \propto M_\star^{-1/2}$. The diagonal burgundy and green lines represent model targets: a naive naturalness requirement on the coupling with cutoff scale $\L =3$ TeV, and the boundary of physically-realized Higgs Portal models utilizing a relaxion, respectively \cite{Flacke:2016szy,Choi:2016luu}. We observe that the relaxion benchmark is reachable by future space-based clocks for any of the three couplings we consider in this work, whereas terrestrial clocks may require much greater increases in sensitivity reach to achieve the same for a certain mass range. 

A space probe with a nuclear clock would allow one to probe a vastly larger parameter space, reaching for the first time physically-motivated model space for Higgs-relaxion mixing (below the green line) for both photon and QCD couplings.

\section{Spatial Variation of Fundamental Constants}

With our proposal of a space mission with a clock-comparison experiment in an inner solar orbit, one can also test the variations of fundamental constants due to the change in the gravitational potential
during the satellite transit to its orbit. Such new physics is usually parameterized as \cite{Safronova:2017xyt,Lange:2020cul}
\begin{equation} \label{eq:dX}
 k_X \equiv c^2\frac{\d X}{X\,\d U}.
\end{equation}
We quantify the change in gravitational potential as $\d U$ between the positions of two clock measurements, and $X=\alpha$, $\mu$, or $m_q/\Lambda_{QCD}$.
There are essentially differential redshift experiments, referred to as “null” experiments in \cite{Will2014}.
Monitoring ratio of clocks as the satellite moves deeper in the solar system can set strong constraints on the  parameters $k_X$, as
$\left(k_X\right)_{\rm exp} = \left(\d X/X\right)_{\rm exp}c^2/\d U$.

Previous studies utilize the seasonal variation in Earth's orbital distance to the Sun, which gives rise to a difference of $\d U/c^2 \simeq 3.3\times10^{-10}$, which is used to constraint $k_X$ \cite{Lange:2020cul}.
For a probe at $0.1$ AU, as we have considered in this proposal,
one can expect the change of the potential in comparison with 1 AU of $\delta U/c^2\sim 9\times10^{-8}$,  about $300$ times larger than that of the Earth's annual modulation. If the same uncertainty on measuring $\delta X/X$ can be reached in space as on Earth, one can therefore achieve constraints on $k_X$ that are a factor of $\sim 300$ stronger, barring systematic uncertainties due to the space mission. 

Our present proposal does not require an optical link enabling comparing the satellite and Earth-based clocks. If such a link can be achieved, one can also directly test general relativity and provide a direct bound on the anomalous gravitational redshift exceeding present bounds by orders of magnitude \cite{Will2014,Sch09,Lit2021}.

\section{Other Applications And Outlook}

We present an exciting opportunity to study ultralight dark matter in unexplored and theoretically motivated regions with atomic, molecular, and nuclear clocks in space. Such clocks can probe a very large parameter space for ULDM bound to the Sun, with the possibility in the near future of reaching well-motivated theory targets. Additionally, a clock in near approach to the Sun can significantly improve limits on spatial variation of fundamental constants.

Below, we briefly discuss some natural extensions of our ideas, as well as other well-motivated physics topics that space and quantum technologies can probe.

{\it Space Quantum Clock Networks (SQCN) --}
A network of clocks in space and on Earth can study many fundamental physics topics, including transient topological dark matter \cite{Derevianko:2013oaa,Wcislo:2018ojh}, and multimessenger signatures of exotic particles \cite{Derevianko:2021wgw}.
In our consideration, if a signal were to be present, the comparison of ground- and space-based clocks could help to map the density of DM in the vicinity of Earth to further constrain the bound DM scenario. One could set up a network of atomic and nuclear clocks on Earth and in space for this purpose. A high-precision clock in space with an optical link to Earth will also enable us to compare optical clocks in any place on Earth \cite{Shen:2021scn}, without the need for a fiber-link connection \cite{Gozzard:2021wkf}.

{\it  Screening --} An additional motivation for a space-based atomic clock arises when the ULDM scalar field possesses quadratic couplings to SM fields. For example, in the presence of a coupling of the form $\Lcal \supset g_2\,\frac{\phi^2}{M_P^2}\,m_i\,\bar{\psi_i}\psi_i,$ (where $\psi_i$ are SM fields of mass $m_i$) with a positive coefficient, there is a screening\footnote{If the quadratic coupling has a negative sign, the effect is instead an anti-screening of the field \cite{Hees:2018fpg}.} of the field value in the vicinity of the Earth, due to a backreaction of the large number density $\bar{\psi}\psi$ of, e.g., electrons or neutrons in the Earth, rapidly reducing the sensitivity of terrestrial experiments  \cite{Hees:2018fpg}; the effect is even more severe for transients. A space-based probe considered in this work would not be subject to this Earth screening effect.

\section*{Acknowledgement}

We thank Kevork Abazajian, Asantha Cooray, ChunChia Chen, Jonathan Feng, Manoj Kaplinghat, Hyungjin Kim, David Leibrandt, Gilad Perez, Stefano Profumo, and Tim Tait for useful discussions.

A part of this work was performed at the Aspen Center for Physics, which is supported by the National Science Foundation grant PHY-1607611.
JE thanks the Galileo Galilei Institute for Theoretical Physics for the hospitality and the INFN for partial support during the completion of this work.
The work of Y-DT is supported in part by U.S. National Science Foundation Grant No. PHY-1915005.
A part of this document was prepared by Y-DT using the resources of the Fermi National Accelerator Laboratory (Fermilab), a U.S. Department of Energy, Office of Science, HEP User Facility. Fermilab is managed by Fermi Research Alliance, LLC (FRA), acting under Contract No. DE-AC02-07CH11359.
The work of JE was supported by the World Premier International Research Center Initiative (WPI), MEXT, Japan, and by the JSPS KAKENHI Grant Numbers 21H05451 and 21K20366. This work is supported in part by US NSF Grants No. PHY-2012068 and OMA-2016244.
This work is a part of the “Thorium Nuclear Clock” project that has received funding from the European Research Council (ERC) under the European Union’s Horizon 2020 research and innovation program (Grant Agreement No. 856415).

\bibliographystyle{JHEP}
\bibliography{ULDM_SpaceClocks,clocks}

\providecommand{\href}[2]{#2}\begingroup\raggedright\begin{thebibliography}{100}

\bibitem{Graham:2015cka}
P.W.~Graham, D.E.~Kaplan and S.~Rajendran, \emph{{Cosmological Relaxation of
  the Electroweak Scale}},
  \href{https://doi.org/10.1103/PhysRevLett.115.221801}{\emph{Phys. Rev. Lett.}
  {\bfseries 115} (2015) 221801}
  [\href{https://arxiv.org/abs/1504.07551}{{\ttfamily 1504.07551}}].

\bibitem{Banerjee:2018xmn}
A.~Banerjee, H.~Kim and G.~Perez, \emph{{Coherent relaxion dark matter}},
  \href{https://doi.org/10.1103/PhysRevD.100.115026}{\emph{Phys. Rev. D}
  {\bfseries 100} (2019) 115026}
  [\href{https://arxiv.org/abs/1810.01889}{{\ttfamily 1810.01889}}].

\bibitem{Svrcek:2006yi}
P.~Svrcek and E.~Witten, \emph{{Axions In String Theory}},
  \href{https://doi.org/10.1088/1126-6708/2006/06/051}{\emph{JHEP} {\bfseries
  06} (2006) 051} [\href{https://arxiv.org/abs/hep-th/0605206}{{\ttfamily
  hep-th/0605206}}].

\bibitem{Arvanitaki:2009fg}
A.~Arvanitaki, S.~Dimopoulos, S.~Dubovsky, N.~Kaloper and J.~March-Russell,
  \emph{{String Axiverse}},
  \href{https://doi.org/10.1103/PhysRevD.81.123530}{\emph{Phys. Rev. D}
  {\bfseries 81} (2010) 123530}
  [\href{https://arxiv.org/abs/0905.4720}{{\ttfamily 0905.4720}}].

\bibitem{Cicoli:2012sz}
M.~Cicoli, M.~Goodsell and A.~Ringwald, \emph{{The type IIB string axiverse and
  its low-energy phenomenology}},
  \href{https://doi.org/10.1007/JHEP10(2012)146}{\emph{JHEP} {\bfseries 10}
  (2012) 146} [\href{https://arxiv.org/abs/1206.0819}{{\ttfamily 1206.0819}}].

\bibitem{Visinelli:2018utg}
L.~Visinelli and S.~Vagnozzi, \emph{{Cosmological window onto the string
  axiverse and the supersymmetry breaking scale}},
  \href{https://doi.org/10.1103/PhysRevD.99.063517}{\emph{Phys. Rev. D}
  {\bfseries 99} (2019) 063517}
  [\href{https://arxiv.org/abs/1809.06382}{{\ttfamily 1809.06382}}].

\bibitem{Peccei:1987mm}
R.D.~Peccei, J.~Sola and C.~Wetterich, \emph{{Adjusting the Cosmological
  Constant Dynamically: Cosmons and a New Force Weaker Than Gravity}},
  \href{https://doi.org/10.1016/0370-2693(87)91191-9}{\emph{Phys. Lett. B}
  {\bfseries 195} (1987) 183}.

\bibitem{Wetterich:1987fm}
C.~Wetterich, \emph{{Cosmology and the Fate of Dilatation Symmetry}},
  \href{https://doi.org/10.1016/0550-3213(88)90193-9}{\emph{Nucl. Phys. B}
  {\bfseries 302} (1988) 668}
  [\href{https://arxiv.org/abs/1711.03844}{{\ttfamily 1711.03844}}].

\bibitem{Ratra:1987rm}
B.~Ratra and P.J.E.~Peebles, \emph{{Cosmological Consequences of a Rolling
  Homogeneous Scalar Field}},
  \href{https://doi.org/10.1103/PhysRevD.37.3406}{\emph{Phys. Rev. D}
  {\bfseries 37} (1988) 3406}.

\bibitem{Wetterich:2002ic}
C.~Wetterich, \emph{{Probing quintessence with time variation of couplings}},
  \href{https://doi.org/10.1088/1475-7516/2003/10/002}{\emph{JCAP} {\bfseries
  10} (2003) 002} [\href{https://arxiv.org/abs/hep-ph/0203266}{{\ttfamily
  hep-ph/0203266}}].

\bibitem{Khoury:2003rn}
J.~Khoury and A.~Weltman, \emph{{Chameleon cosmology}},
  \href{https://doi.org/10.1103/PhysRevD.69.044026}{\emph{Phys. Rev. D}
  {\bfseries 69} (2004) 044026}
  [\href{https://arxiv.org/abs/astro-ph/0309411}{{\ttfamily
  astro-ph/0309411}}].

\bibitem{Kim:2002tq}
J.E.~Kim and H.P.~Nilles, \emph{{A Quintessential axion}},
  \href{https://doi.org/10.1016/S0370-2693(02)03148-9}{\emph{Phys. Lett. B}
  {\bfseries 553} (2003) 1}
  [\href{https://arxiv.org/abs/hep-ph/0210402}{{\ttfamily hep-ph/0210402}}].

\bibitem{Ibe:2018ffn}
M.~Ibe, M.~Yamazaki and T.T.~Yanagida, \emph{{Quintessence Axion Revisited in
  Light of Swampland Conjectures}},
  \href{https://doi.org/10.1088/1361-6382/ab5197}{\emph{Class. Quant. Grav.}
  {\bfseries 36} (2019) 235020}
  [\href{https://arxiv.org/abs/1811.04664}{{\ttfamily 1811.04664}}].

\bibitem{Choi:2021aze}
G.~Choi, W.~Lin, L.~Visinelli and T.T.~Yanagida, \emph{{Cosmic Birefringence
  and Electroweak Axion Dark Energy}},
  \href{https://arxiv.org/abs/2106.12602}{{\ttfamily 2106.12602}}.

\bibitem{Hu:2000ke}
W.~Hu, R.~Barkana and A.~Gruzinov, \emph{{Cold and fuzzy dark matter}},
  \href{https://doi.org/10.1103/PhysRevLett.85.1158}{\emph{Phys. Rev. Lett.}
  {\bfseries 85} (2000) 1158}
  [\href{https://arxiv.org/abs/astro-ph/0003365}{{\ttfamily
  astro-ph/0003365}}].

\bibitem{Hui:2016ltb}
L.~Hui, J.P.~Ostriker, S.~Tremaine and E.~Witten, \emph{{Ultralight scalars as
  cosmological dark matter}},
  \href{https://doi.org/10.1103/PhysRevD.95.043541}{\emph{Phys. Rev. D}
  {\bfseries 95} (2017) 043541}
  [\href{https://arxiv.org/abs/1610.08297}{{\ttfamily 1610.08297}}].

\bibitem{Mocz:2019pyf}
P.~Mocz et~al., \emph{{First star-forming structures in fuzzy cosmic
  filaments}},
  \href{https://doi.org/10.1103/PhysRevLett.123.141301}{\emph{Phys. Rev. Lett.}
  {\bfseries 123} (2019) 141301}
  [\href{https://arxiv.org/abs/1910.01653}{{\ttfamily 1910.01653}}].

\bibitem{PhysRevD.50.3614}
Y.~Su, B.R.~Heckel, E.G.~Adelberger, J.H.~Gundlach, M.~Harris, G.L.~Smith
  et~al., \emph{New tests of the universality of free fall},
  \href{https://doi.org/10.1103/PhysRevD.50.3614}{\emph{Phys. Rev. D}
  {\bfseries 50} (1994) 3614}.

\bibitem{Hoyle:2004cw}
C.D.~Hoyle, D.J.~Kapner, B.R.~Heckel, E.G.~Adelberger, J.H.~Gundlach,
  U.~Schmidt et~al., \emph{{Sub-millimeter tests of the gravitational
  inverse-square law}},
  \href{https://doi.org/10.1103/PhysRevD.70.042004}{\emph{Phys. Rev. D}
  {\bfseries 70} (2004) 042004}
  [\href{https://arxiv.org/abs/hep-ph/0405262}{{\ttfamily hep-ph/0405262}}].

\bibitem{Williams:2004qba}
J.G.~Williams, S.G.~Turyshev and D.H.~Boggs, \emph{{Progress in lunar laser
  ranging tests of relativistic gravity}},
  \href{https://doi.org/10.1103/PhysRevLett.93.261101}{\emph{Phys. Rev. Lett.}
  {\bfseries 93} (2004) 261101}
  [\href{https://arxiv.org/abs/gr-qc/0411113}{{\ttfamily gr-qc/0411113}}].

\bibitem{Mota:2006fz}
D.F.~Mota and D.J.~Shaw, \emph{{Evading Equivalence Principle Violations,
  Cosmological and other Experimental Constraints in Scalar Field Theories with
  a Strong Coupling to Matter}},
  \href{https://doi.org/10.1103/PhysRevD.75.063501}{\emph{Phys. Rev. D}
  {\bfseries 75} (2007) 063501}
  [\href{https://arxiv.org/abs/hep-ph/0608078}{{\ttfamily hep-ph/0608078}}].

\bibitem{Brax:2007vm}
P.~Brax, C.~van~de Bruck, A.-C.~Davis, D.F.~Mota and D.J.~Shaw,
  \emph{{Detecting chameleons through Casimir force measurements}},
  \href{https://doi.org/10.1103/PhysRevD.76.124034}{\emph{Phys. Rev. D}
  {\bfseries 76} (2007) 124034}
  [\href{https://arxiv.org/abs/0709.2075}{{\ttfamily 0709.2075}}].

\bibitem{Schlamminger:2007ht}
S.~Schlamminger, K.Y.~Choi, T.A.~Wagner, J.H.~Gundlach and E.G.~Adelberger,
  \emph{{Test of the equivalence principle using a rotating torsion balance}},
  \href{https://doi.org/10.1103/PhysRevLett.100.041101}{\emph{Phys. Rev. Lett.}
  {\bfseries 100} (2008) 041101}
  [\href{https://arxiv.org/abs/0712.0607}{{\ttfamily 0712.0607}}].

\bibitem{Brax:2011hb}
P.~Brax and G.~Pignol, \emph{{Strongly Coupled Chameleons and the Neutronic
  Quantum Bouncer}},
  \href{https://doi.org/10.1103/PhysRevLett.107.111301}{\emph{Phys. Rev. Lett.}
  {\bfseries 107} (2011) 111301}
  [\href{https://arxiv.org/abs/1105.3420}{{\ttfamily 1105.3420}}].

\bibitem{Wagner:2012ui}
T.A.~Wagner, S.~Schlamminger, J.H.~Gundlach and E.G.~Adelberger,
  \emph{{Torsion-balance tests of the weak equivalence principle}},
  \href{https://doi.org/10.1088/0264-9381/29/18/184002}{\emph{Class. Quant.
  Grav.} {\bfseries 29} (2012) 184002}
  [\href{https://arxiv.org/abs/1207.2442}{{\ttfamily 1207.2442}}].

\bibitem{Burrage:2014oza}
C.~Burrage, E.J.~Copeland and E.A.~Hinds, \emph{{Probing Dark Energy with Atom
  Interferometry}},
  \href{https://doi.org/10.1088/1475-7516/2015/03/042}{\emph{JCAP} {\bfseries
  03} (2015) 042} [\href{https://arxiv.org/abs/1408.1409}{{\ttfamily
  1408.1409}}].

\bibitem{Foot:2014osa}
R.~Foot and S.~Vagnozzi, \emph{{Diurnal modulation signal from dissipative
  hidden sector dark matter}},
  \href{https://doi.org/10.1016/j.physletb.2015.06.063}{\emph{Phys. Lett. B}
  {\bfseries 748} (2015) 61} [\href{https://arxiv.org/abs/1412.0762}{{\ttfamily
  1412.0762}}].

\bibitem{Roberts:2017hla}
B.M.~Roberts, G.~Blewitt, C.~Dailey, M.~Murphy, M.~Pospelov, A.~Rollings
  et~al., \emph{{Search for domain wall dark matter with atomic clocks on board
  global positioning system satellites}},
  \href{https://doi.org/10.1038/s41467-017-01440-4}{\emph{Nature Commun.}
  {\bfseries 8} (2017) 1195}
  [\href{https://arxiv.org/abs/1704.06844}{{\ttfamily 1704.06844}}].

\bibitem{Jain:2012tn}
B.~Jain, V.~Vikram and J.~Sakstein, \emph{{Astrophysical Tests of Modified
  Gravity: Constraints from Distance Indicators in the Nearby Universe}},
  \href{https://doi.org/10.1088/0004-637X/779/1/39}{\emph{Astrophys. J.}
  {\bfseries 779} (2013) 39} [\href{https://arxiv.org/abs/1204.6044}{{\ttfamily
  1204.6044}}].

\bibitem{Arvanitaki:2014wva}
A.~Arvanitaki, M.~Baryakhtar and X.~Huang, \emph{{Discovering the QCD Axion
  with Black Holes and Gravitational Waves}},
  \href{https://doi.org/10.1103/PhysRevD.91.084011}{\emph{Phys. Rev. D}
  {\bfseries 91} (2015) 084011}
  [\href{https://arxiv.org/abs/1411.2263}{{\ttfamily 1411.2263}}].

\bibitem{Giannotti:2015kwo}
M.~Giannotti, I.~Irastorza, J.~Redondo and A.~Ringwald, \emph{{Cool WISPs for
  stellar cooling excesses}},
  \href{https://doi.org/10.1088/1475-7516/2016/05/057}{\emph{JCAP} {\bfseries
  05} (2016) 057} [\href{https://arxiv.org/abs/1512.08108}{{\ttfamily
  1512.08108}}].

\bibitem{Foot:2016wvj}
R.~Foot and S.~Vagnozzi, \emph{{Solving the small-scale structure puzzles with
  dissipative dark matter}},
  \href{https://doi.org/10.1088/1475-7516/2016/07/013}{\emph{JCAP} {\bfseries
  07} (2016) 013} [\href{https://arxiv.org/abs/1602.02467}{{\ttfamily
  1602.02467}}].

\bibitem{Caputo:2017zqh}
A.~Caputo, J.~Zavala and D.~Blas, \emph{{Binary pulsars as probes of a Galactic
  dark matter disk}},
  \href{https://doi.org/10.1016/j.dark.2017.10.005}{\emph{Phys. Dark Univ.}
  {\bfseries 19} (2018) 1} [\href{https://arxiv.org/abs/1709.03991}{{\ttfamily
  1709.03991}}].

\bibitem{Baryakhtar:2017ngi}
M.~Baryakhtar, R.~Lasenby and M.~Teo, \emph{{Black Hole Superradiance
  Signatures of Ultralight Vectors}},
  \href{https://doi.org/10.1103/PhysRevD.96.035019}{\emph{Phys. Rev. D}
  {\bfseries 96} (2017) 035019}
  [\href{https://arxiv.org/abs/1704.05081}{{\ttfamily 1704.05081}}].

\bibitem{Croon:2020oga}
D.~Croon, S.D.~McDermott and J.~Sakstein, \emph{{New physics and the black hole
  mass gap}}, \href{https://doi.org/10.1103/PhysRevD.102.115024}{\emph{Phys.
  Rev. D} {\bfseries 102} (2020) 115024}
  [\href{https://arxiv.org/abs/2007.07889}{{\ttfamily 2007.07889}}].

\bibitem{Hlozek:2014lca}
R.~Hlozek, D.~Grin, D.J.E.~Marsh and P.G.~Ferreira, \emph{{A search for
  ultralight axions using precision cosmological data}},
  \href{https://doi.org/10.1103/PhysRevD.91.103512}{\emph{Phys. Rev. D}
  {\bfseries 91} (2015) 103512}
  [\href{https://arxiv.org/abs/1410.2896}{{\ttfamily 1410.2896}}].

\bibitem{Baumann:2015rya}
D.~Baumann, D.~Green, J.~Meyers and B.~Wallisch, \emph{{Phases of New Physics
  in the CMB}},
  \href{https://doi.org/10.1088/1475-7516/2016/01/007}{\emph{JCAP} {\bfseries
  01} (2016) 007} [\href{https://arxiv.org/abs/1508.06342}{{\ttfamily
  1508.06342}}].

\bibitem{DEramo:2018vss}
F.~D'Eramo, R.Z.~Ferreira, A.~Notari and J.L.~Bernal, \emph{{Hot Axions and the
  $H_0$ tension}},
  \href{https://doi.org/10.1088/1475-7516/2018/11/014}{\emph{JCAP} {\bfseries
  11} (2018) 014} [\href{https://arxiv.org/abs/1808.07430}{{\ttfamily
  1808.07430}}].

\bibitem{Ade:2018sbj}
{\scshape Simons Observatory} collaboration, \emph{{The Simons Observatory:
  Science goals and forecasts}},
  \href{https://doi.org/10.1088/1475-7516/2019/02/056}{\emph{JCAP} {\bfseries
  02} (2019) 056} [\href{https://arxiv.org/abs/1808.07445}{{\ttfamily
  1808.07445}}].

\bibitem{Vagnozzi:2019ezj}
S.~Vagnozzi, \emph{{New physics in light of the $H_0$ tension: An alternative
  view}}, \href{https://doi.org/10.1103/PhysRevD.102.023518}{\emph{Phys. Rev.
  D} {\bfseries 102} (2020) 023518}
  [\href{https://arxiv.org/abs/1907.07569}{{\ttfamily 1907.07569}}].

\bibitem{Pitjev:2013sfa}
N.P.~Pitjev and E.V.~Pitjeva, \emph{{Constraints on dark matter in the solar
  system}}, \href{https://doi.org/10.1134/S1063773713020060}{\emph{Astron.
  Lett.} {\bfseries 39} (2013) 141}
  [\href{https://arxiv.org/abs/1306.5534}{{\ttfamily 1306.5534}}].

\bibitem{Tsai:2021irw}
Y.-D.~Tsai, Y.~Wu, S.~Vagnozzi and L.~Visinelli, \emph{{Asteroid astrometry as
  a fifth-force and ultralight dark sector probe}},
  \href{https://arxiv.org/abs/2107.04038}{{\ttfamily 2107.04038}}.

\bibitem{Poddar:2020exe}
T.~Kumar~Poddar, S.~Mohanty and S.~Jana, \emph{{Constraints on long range force
  from perihelion precession of planets in a gauged $L_e-L_{\mu,\tau}$
  scenario}},  \href{https://arxiv.org/abs/2002.02935}{{\ttfamily 2002.02935}}.

\bibitem{Poddar:2021ose}
T.K.~Poddar, \emph{{Constraints on ultralight axions, vector gauge bosons, and
  unparticles from geodetic and frame-dragging effects}},
  \href{https://arxiv.org/abs/2111.05632}{{\ttfamily 2111.05632}}.

\bibitem{Puetzfeld:2019kki}
D.~Puetzfeld and C.~L\"ammerzahl, eds., \emph{{Relativistic Geodesy}},
  vol.~196, Springer (2019),
  \href{https://doi.org/10.1007/978-3-030-11500-5}{10.1007/978-3-030-11500-5}.

\bibitem{Gozzard:2021wkf}
D.R.~Gozzard, L.A.~Howard, B.P.~Dix-Matthews, S.~Karpathakis, C.~Gravestock and
  S.W.~Schediwy, \emph{{Ultra-stable Free-Space Laser Links for a Global
  Network of Optical Atomic Clocks}},
  \href{https://arxiv.org/abs/2103.12909}{{\ttfamily 2103.12909}}.

\bibitem{Yin17}
J.~{Yin}, Y.~{Cao}, Y.-H.~{Li}, S.-K.~{Liao}, L.~{Zhang}, J.-G.~{Ren} et~al.,
  \emph{{Satellite-Based Entanglement Distribution Over 1200 kilometers}},
  \href{https://doi.org/10.1126/science.aan3211}{\emph{Science} {\bfseries 356}
  (2017) 1140}.

\bibitem{Burt2021}
E.~Burt, J.~Prestage, R.~Tjoelker, D.~Enzer, D.~Kuang, D.~Murphy et~al.,
  \emph{Demonstration of a trapped-ion atomic clock in space},
  \href{https://doi.org/10.1038/s41586-021-03571-7}{\emph{Nature} {\bfseries
  595} (2021) 43}.

\bibitem{CAC2018}
L.~Liu, D.-S.~L{\"u}, W.~biao Chen, T.~Li, Q.~Qu, B.~Wang et~al.,
  \emph{In-orbit operation of an atomic clock based on laser-cooled 87rb
  atoms}, {\emph{Nature Communications} {\bfseries 2760} (2018) }.

\bibitem{Arvanitaki:2014faa}
A.~Arvanitaki, J.~Huang and K.~Van~Tilburg, \emph{{Searching for dilaton dark
  matter with atomic clocks}},
  \href{https://doi.org/10.1103/PhysRevD.91.015015}{\emph{Phys. Rev. D}
  {\bfseries 91} (2015) 015015}
  [\href{https://arxiv.org/abs/1405.2925}{{\ttfamily 1405.2925}}].

\bibitem{VanTilburg:2015oza}
K.~Van~Tilburg, N.~Leefer, L.~Bougas and D.~Budker, \emph{{Search for
  ultralight scalar dark matter with atomic spectroscopy}},
  \href{https://doi.org/10.1103/PhysRevLett.115.011802}{\emph{Phys. Rev. Lett.}
  {\bfseries 115} (2015) 011802}
  [\href{https://arxiv.org/abs/1503.06886}{{\ttfamily 1503.06886}}].

\bibitem{Banerjee:2019epw}
A.~Banerjee, D.~Budker, J.~Eby, H.~Kim and G.~Perez, \emph{{Relaxion Stars and
  their detection via Atomic Physics}},
  \href{https://doi.org/10.1038/s42005-019-0260-3}{\emph{Commun. Phys.}
  {\bfseries 3} (2020) 1} [\href{https://arxiv.org/abs/1902.08212}{{\ttfamily
  1902.08212}}].

\bibitem{Banerjee:2019xuy}
A.~Banerjee, D.~Budker, J.~Eby, V.V.~Flambaum, H.~Kim, O.~Matsedonskyi et~al.,
  \emph{{Searching for Earth/Solar Axion Halos}},
  \href{https://doi.org/10.1007/JHEP09(2020)004}{\emph{JHEP} {\bfseries 09}
  (2020) 004} [\href{https://arxiv.org/abs/1912.04295}{{\ttfamily
  1912.04295}}].

\bibitem{Safronova:2017xyt}
M.S.~Safronova, D.~Budker, D.~DeMille, D.F.J.~Kimball, A.~Derevianko and
  C.W.~Clark, \emph{{Search for New Physics with Atoms and Molecules}},
  \href{https://doi.org/10.1103/RevModPhys.90.025008}{\emph{Rev. Mod. Phys.}
  {\bfseries 90} (2018) 025008}
  [\href{https://arxiv.org/abs/1710.01833}{{\ttfamily 1710.01833}}].

\bibitem{Lange:2020cul}
R.~Lange, N.~Huntemann, J.M.~Rahm, C.~Sanner, H.~Shao, B.~Lipphardt et~al.,
  \emph{{Improved limits for violations of local position invariance from
  atomic clock comparisons}},
  \href{https://doi.org/10.1103/PhysRevLett.126.011102}{\emph{Phys. Rev. Lett.}
  {\bfseries 126} (2021) 011102}
  [\href{https://arxiv.org/abs/2010.06620}{{\ttfamily 2010.06620}}].

\bibitem{Damour:2010rm}
T.~Damour and J.F.~Donoghue, \emph{{Phenomenology of the Equivalence Principle
  with Light Scalars}},
  \href{https://doi.org/10.1088/0264-9381/27/20/202001}{\emph{Class. Quant.
  Grav.} {\bfseries 27} (2010) 202001}
  [\href{https://arxiv.org/abs/1007.2790}{{\ttfamily 1007.2790}}].

\bibitem{Damour:2010rp}
T.~Damour and J.F.~Donoghue, \emph{{Equivalence Principle Violations and
  Couplings of a Light Dilaton}},
  \href{https://doi.org/10.1103/PhysRevD.82.084033}{\emph{Phys. Rev. D}
  {\bfseries 82} (2010) 084033}
  [\href{https://arxiv.org/abs/1007.2792}{{\ttfamily 1007.2792}}].

\bibitem{Hees:2016gop}
A.~Hees, J.~Gu\'ena, M.~Abgrall, S.~Bize and P.~Wolf, \emph{{Searching for an
  oscillating massive scalar field as a dark matter candidate using atomic
  hyperfine frequency comparisons}},
  \href{https://doi.org/10.1103/PhysRevLett.117.061301}{\emph{Phys. Rev. Lett.}
  {\bfseries 117} (2016) 061301}
  [\href{https://arxiv.org/abs/1604.08514}{{\ttfamily 1604.08514}}].

\bibitem{Wcislo2018}
P.~Wcislo, P.~Ablewski, K.~Beloy, S.~Bilicki, M.~Bober, R.~Brown et~al.,
  \emph{New bounds on dark matter coupling from a global network of optical
  atomic clocks}, \href{https://doi.org/10.1126/sciadv.aau4869}{\emph{Science
  Advances} {\bfseries 4} (2018) eaau4869}
  [\href{https://arxiv.org/abs/https://www.science.org/doi/pdf/10.1126/sciadv.aau4869}{{\ttfamily
  https://www.science.org/doi/pdf/10.1126/sciadv.aau4869}}].

\bibitem{Aharony:2019iad}
S.~Aharony, N.~Akerman, R.~Ozeri, G.~Perez, I.~Savoray and R.~Shaniv,
  \emph{{Constraining Rapidly Oscillating Scalar Dark Matter Using Dynamic
  Decoupling}}, \href{https://doi.org/10.1103/PhysRevD.103.075017}{\emph{Phys.
  Rev. D} {\bfseries 103} (2021) 075017}
  [\href{https://arxiv.org/abs/1902.02788}{{\ttfamily 1902.02788}}].

\bibitem{Antypas:2019qji}
D.~Antypas, O.~Tretiak, A.~Garcon, R.~Ozeri, G.~Perez and D.~Budker,
  \emph{{Scalar dark matter in the radio-frequency band: atomic-spectroscopy
  search results}},
  \href{https://doi.org/10.1103/PhysRevLett.123.141102}{\emph{Phys. Rev. Lett.}
  {\bfseries 123} (2019) 141102}
  [\href{https://arxiv.org/abs/1905.02968}{{\ttfamily 1905.02968}}].

\bibitem{Kennedy:2020bac}
C.J.~Kennedy, E.~Oelker, J.M.~Robinson, T.~Bothwell, D.~Kedar, W.R.~Milner
  et~al., \emph{{Precision Metrology Meets Cosmology: Improved Constraints on
  Ultralight Dark Matter from Atom-Cavity Frequency Comparisons}},
  \href{https://doi.org/10.1103/PhysRevLett.125.201302}{\emph{Phys. Rev. Lett.}
  {\bfseries 125} (2020) 201302}
  [\href{https://arxiv.org/abs/2008.08773}{{\ttfamily 2008.08773}}].

\bibitem{Savalle:2020vgz}
E.~Savalle, A.~Hees, F.~Frank, E.~Cantin, P.-E.~Pottie, B.M.~Roberts et~al.,
  \emph{{Searching for Dark Matter with an Optical Cavity and an Unequal-Delay
  Interferometer}},
  \href{https://doi.org/10.1103/PhysRevLett.126.051301}{\emph{Phys. Rev. Lett.}
  {\bfseries 126} (2021) 051301}
  [\href{https://arxiv.org/abs/2006.07055}{{\ttfamily 2006.07055}}].

\bibitem{Oswald:2021vtc}
R.~Oswald et~al., \emph{{Search for oscillations of fundamental constants using
  molecular spectroscopy}},  \href{https://arxiv.org/abs/2111.06883}{{\ttfamily
  2111.06883}}.

\bibitem{Berge:2017ovy}
J.~Berg\'e, P.~Brax, G.~M\'etris, M.~Pernot-Borr\`as, P.~Touboul and
  J.-P.~Uzan, \emph{{MICROSCOPE Mission: First Constraints on the Violation of
  the Weak Equivalence Principle by a Light Scalar Dilaton}},
  \href{https://doi.org/10.1103/PhysRevLett.120.141101}{\emph{Phys. Rev. Lett.}
  {\bfseries 120} (2018) 141101}
  [\href{https://arxiv.org/abs/1712.00483}{{\ttfamily 1712.00483}}].

\bibitem{Hees:2018fpg}
A.~Hees, O.~Minazzoli, E.~Savalle, Y.V.~Stadnik and P.~Wolf, \emph{{Violation
  of the equivalence principle from light scalar dark matter}},
  \href{https://doi.org/10.1103/PhysRevD.98.064051}{\emph{Phys. Rev. D}
  {\bfseries 98} (2018) 064051}
  [\href{https://arxiv.org/abs/1807.04512}{{\ttfamily 1807.04512}}].

\bibitem{Anderson:2020rdk}
N.B.~Anderson, A.~Partenheimer and T.D.~Wiser, \emph{{Direct detection
  signatures of a primordial Solar dark matter halo}},
  \href{https://arxiv.org/abs/2007.11016}{{\ttfamily 2007.11016}}.

\bibitem{VanTilburg:2020jvl}
K.~Van~Tilburg, \emph{{Stellar Basins of Gravitationally Bound Particles}},
  \href{https://arxiv.org/abs/2006.12431}{{\ttfamily 2006.12431}}.

\bibitem{Lasenby:2020goo}
R.~Lasenby and K.~Van~Tilburg, \emph{{Dark photons in the solar basin}},
  \href{https://doi.org/10.1103/PhysRevD.104.023020}{\emph{Phys. Rev. D}
  {\bfseries 104} (2021) 023020}
  [\href{https://arxiv.org/abs/2008.08594}{{\ttfamily 2008.08594}}].

\bibitem{Veltmaat:2019hou}
J.~Veltmaat, B.~Schwabe and J.C.~Niemeyer, \emph{{Baryon-driven growth of
  solitonic cores in fuzzy dark matter halos}},
  \href{https://doi.org/10.1103/PhysRevD.101.083518}{\emph{Phys. Rev. D}
  {\bfseries 101} (2020) 083518}
  [\href{https://arxiv.org/abs/1911.09614}{{\ttfamily 1911.09614}}].

\bibitem{Kaup:1968zz}
D.J.~Kaup, \emph{{Klein-Gordon Geon}},
  \href{https://doi.org/10.1103/PhysRev.172.1331}{\emph{Phys. Rev.} {\bfseries
  172} (1968) 1331}.

\bibitem{Ruffini:1969qy}
R.~Ruffini and S.~Bonazzola, \emph{{Systems of selfgravitating particles in
  general relativity and the concept of an equation of state}},
  \href{https://doi.org/10.1103/PhysRev.187.1767}{\emph{Phys. Rev.} {\bfseries
  187} (1969) 1767}.

\bibitem{Colpi:1986ye}
M.~Colpi, S.L.~Shapiro and I.~Wasserman, \emph{{Boson Stars: Gravitational
  Equilibria of Selfinteracting Scalar Fields}},
  \href{https://doi.org/10.1103/PhysRevLett.57.2485}{\emph{Phys. Rev. Lett.}
  {\bfseries 57} (1986) 2485}.

\bibitem{Chavanis:2011zi}
P.-H.~Chavanis, \emph{{Mass-radius relation of Newtonian self-gravitating
  Bose-Einstein condensates with short-range interactions: I. Analytical
  results}}, \href{https://doi.org/10.1103/PhysRevD.84.043531}{\emph{Phys.
  Rev.} {\bfseries D84} (2011) 043531}
  [\href{https://arxiv.org/abs/1103.2050}{{\ttfamily 1103.2050}}].

\bibitem{Der18}
A.~Derevianko, \emph{Detecting dark-matter waves with a network of
  precision-measurement tools},
  \href{https://doi.org/10.1103/physreva.97.042506}{\emph{Physical Review A}
  {\bfseries 97} (2018) }.

\bibitem{CenBlaCon20}
G.P.~Centers, J.W.~Blanchard, J.~Conrad, N.L.~Figueroa, A.~Garcon,
  A.V.~Gramolin et~al., \emph{Stochastic fluctuations of bosonic dark matter},
  2020.

\bibitem{Wcislo2016}
P.~Wcislo, P.~Morzy{\'n}ski, M.~Bober, A.~Cygan, D.~Lisak, R.~Ciurylo et~al.,
  \emph{Experimental constraint on dark matter detection with optical atomic
  clocks}, \href{https://doi.org/10.1038/s41550-016-0009}{\emph{Nature
  Astronomy} {\bfseries 1} (2016) 0009}.

\bibitem{FlaDzu09}
V.V.~{Flambaum} and V.A.~{Dzuba}, \emph{{Search for variation of the
  fundamental constants in atomic, molecular, and nuclear spectra}},
  \href{https://doi.org/10.1139/P08-072}{\emph{Canadian Journal of Physics}
  {\bfseries 87} (2009) 25} [\href{https://arxiv.org/abs/0805.0462}{{\ttfamily
  0805.0462}}].

\bibitem{LudBoyYe15}
A.D.~{Ludlow}, M.M.~{Boyd}, J.~{Ye}, E.~{Peik} and P.O.~{Schmidt},
  \emph{{Optical atomic clocks}},
  \href{https://doi.org/10.1103/RevModPhys.87.637}{\emph{Reviews of Modern
  Physics} {\bfseries 87} (2015) 637}
  [\href{https://arxiv.org/abs/1407.3493}{{\ttfamily 1407.3493}}].

\bibitem{LanHunRah21}
R.~{Lange}, N.~{Huntemann}, J.M.~{Rahm}, C.~{Sanner}, H.~{Shao}, B.~{Lipphardt}
  et~al., \emph{{Improved Limits for Violations of Local Position Invariance
  from Atomic Clock Comparisons}},
  \href{https://doi.org/10.1103/PhysRevLett.126.011102}{\emph{\prl} {\bfseries
  126} (2021) 011102} [\href{https://arxiv.org/abs/2010.06620}{{\ttfamily
  2010.06620}}].

\bibitem{PolOatGil13}
N.~{Poli}, C.W.~{Oates}, P.~{Gill} and G.M.~{Tino}, \emph{{Optical atomic
  clocks}}, \href{https://doi.org/10.1393/ncr/i2013-10095-x}{\emph{Nuovo
  Cimento Rivista Serie} {\bfseries 36} (2013) 555}.

\bibitem{BreCheHan19}
S.M.~{Brewer}, J.S.~{Chen}, A.M.~{Hankin}, E.R.~{Clements}, C.W.~{Chou},
  D.J.~{Wineland} et~al., \emph{{$^{27}$Al$^{+}$ Quantum-Logic Clock with a
  Systematic Uncertainty below 10$^{-18}$}},
  \href{https://doi.org/10.1103/PhysRevLett.123.033201}{\emph{\prl} {\bfseries
  123} (2019) 033201} [\href{https://arxiv.org/abs/1902.07694}{{\ttfamily
  1902.07694}}].

\bibitem{SanHunLan19}
C.~{Sanner}, N.~{Huntemann}, R.~{Lange}, C.~{Tamm}, E.~{Peik}, M.S.~{Safronova}
  et~al., \emph{{Optical clock comparison for Lorentz symmetry testing}},
  \href{https://doi.org/10.1038/s41586-019-0972-2}{\emph{\nat} {\bfseries 567}
  (2019) 204} [\href{https://arxiv.org/abs/1809.10742}{{\ttfamily
  1809.10742}}].

\bibitem{BotKedOel19}
T.~Bothwell, D.~Kedar, E.~Oelker, J.M.~Robinson, S.L.~Bromley, W.L.~Tew et~al.,
  \emph{Jila sri optical lattice clock with uncertainty of $2.0 \times
  10^{-18}$}, \href{https://doi.org/10.1088/1681-7575/ab4089}{\emph{Metrologia}
  {\bfseries 56} (2019) 065004}.

\bibitem{WeyGerKaz18}
S.~Weyers, V.~Gerginov, M.~Kazda, J.~Rahm, B.~Lipphardt, G.~Dobrev et~al.,
  \emph{Advances in the accuracy, stability, and reliability of the ptb primary
  fountain clocks},
  \href{https://doi.org/10.1088/1681-7575/aae008}{\emph{Metrologia} {\bfseries
  55} (2018) 789–805}.

\bibitem{BotKenAep21}
T.~Bothwell, C.J.~Kennedy, A.~Aeppli, D.~Kedar, J.M.~Robinson, E.~Oelker
  et~al., \emph{Resolving the gravitational redshift within a millimeter atomic
  sample},  2021.

\bibitem{KelBurKal19}
J.~Stuhler and other, \emph{Opticlock: Transportable and easy-to-operate
  optical single-ion clock},
  \href{https://doi.org/10.1016/j.measen.2021.100264}{\emph{Measurement:
  Sensors} {\bfseries 18} (2021) 100264}.

\bibitem{TakUshOhm20}
U.I.O.N.~Takamoto, M. et~al., \emph{Test of general relativity by a pair of
  transportable optical lattice clocks},
  \href{https://doi.org/10.1038/s41566-020-0619-8}{\emph{Nat. Photonics}
  {\bfseries 14} (2020) 414}.

\bibitem{TanKaeArn19}
T.R.~{Tan}, R.~{Kaewuam}, K.J.~{Arnold}, S.R.~{Chanu}, Z.~{Zhang},
  M.S.~{Safronova} et~al., \emph{{Suppressing Inhomogeneous Broadening in a
  Lutetium Multi-ion Optical Clock}},
  \href{https://doi.org/10.1103/PhysRevLett.123.063201}{\emph{\prl} {\bfseries
  123} (2019) 063201}.

\bibitem{SchBroMcG16}
M.~Schioppo, R.C.~Brown, W.F.~McGrew, N.~Hinkley, R.J.~Fasano, K.~Beloy et~al.,
  \emph{Ultrastable optical clock with two cold-atom ensembles},
  \href{https://doi.org/10.1038/nphoton.2016.231}{\emph{Nature Photonics}
  {\bfseries 11} (2016) 48–52}.

\bibitem{PSP}
NASA, ``Parker solar probe.'' http://parkersolarprobe.jhuapl.edu, 12/9/2021.

\bibitem{FadBerFla20}
P.~Fadeev, J.C.~Berengut and V.V.~Flambaum, \emph{Sensitivity of th229 nuclear
  clock transition to variation of the fine-structure constant},
  \href{https://doi.org/10.1103/physreva.102.052833}{\emph{Physical Review A}
  {\bfseries 102} (2020) }.

\bibitem{Flacke:2016szy}
T.~Flacke, C.~Frugiuele, E.~Fuchs, R.S.~Gupta and G.~Perez,
  \emph{{Phenomenology of relaxion-Higgs mixing}},
  \href{https://doi.org/10.1007/JHEP06(2017)050}{\emph{JHEP} {\bfseries 06}
  (2017) 050} [\href{https://arxiv.org/abs/1610.02025}{{\ttfamily
  1610.02025}}].

\bibitem{Choi:2016luu}
K.~Choi and S.H.~Im, \emph{{Constraints on Relaxion Windows}},
  \href{https://doi.org/10.1007/JHEP12(2016)093}{\emph{JHEP} {\bfseries 12}
  (2016) 093} [\href{https://arxiv.org/abs/1610.00680}{{\ttfamily
  1610.00680}}].

\bibitem{SafBudDem18}
M.S.~{Safronova}, D.~{Budker}, D.~{DeMille}, D.F.J.~{Kimball}, A.~{Derevianko}
  and C.W.~{Clark}, \emph{{Search for new physics with atoms and molecules}},
  \href{https://doi.org/10.1103/RevModPhys.90.025008}{\emph{Reviews of Modern
  Physics} {\bfseries 90} (2018) 025008}
  [\href{https://arxiv.org/abs/1710.01833}{{\ttfamily 1710.01833}}].

\bibitem{HanKuzLun21}
D.~{Hanneke}, B.~{Kuzhan} and A.~{Lunstad}, \emph{{Optical clocks based on
  molecular vibrations as probes of variation of the proton-to-electron mass
  ratio}}, \href{https://doi.org/10.1088/2058-9565/abc863}{\emph{Quantum
  Science and Technology} {\bfseries 6} (2021) 014005}
  [\href{https://arxiv.org/abs/2007.15750}{{\ttfamily 2007.15750}}].

\bibitem{KozSafCre18}
M.G.~{Kozlov}, M.S.~{Safronova}, J.R.~{Crespo L{\'o}pez-Urrutia} and
  P.O.~{Schmidt}, \emph{{Highly charged ions: Optical clocks and applications
  in fundamental physics}},
  \href{https://doi.org/10.1103/RevModPhys.90.045005}{\emph{Reviews of Modern
  Physics} {\bfseries 90} (2018) 045005}
  [\href{https://arxiv.org/abs/1803.06532}{{\ttfamily 1803.06532}}].

\bibitem{PeiSchSaf21}
E.~{Peik}, T.~{Schumm}, M.S.~{Safronova}, A.~{P{\'a}lffy}, J.~{Weitenberg} and
  P.G.~{Thirolf}, \emph{{Nuclear clocks for testing fundamental physics}},
  \href{https://doi.org/10.1088/2058-9565/abe9c2}{\emph{Quantum Science and
  Technology} {\bfseries 6} (2021) 034002}
  [\href{https://arxiv.org/abs/2012.09304}{{\ttfamily 2012.09304}}].

\bibitem{SafPorSan18}
M.S.~{Safronova}, S.G.~{Porsev}, C.~{Sanner} and J.~{Ye}, \emph{{Two Clock
  Transitions in Neutral Yb for the Highest Sensitivity to Variations of the
  Fine-Structure Constant}},
  \href{https://doi.org/10.1103/PhysRevLett.120.173001}{\emph{\prl} {\bfseries
  120} (2018) 173001}.

\bibitem{Fla06}
V.V.~{Flambaum}, \emph{{Enhanced Effect of Temporal Variation of the Fine
  Structure Constant and the Strong Interaction in Th229}},
  \href{https://doi.org/10.1103/PhysRevLett.97.092502}{\emph{\prl} {\bfseries
  97} (2006) 092502} [\href{https://arxiv.org/abs/physics/0601034}{{\ttfamily
  physics/0601034}}].

\bibitem{2007SSRv..131..451C}
J.F.~{Cavanaugh}, J.C.~{Smith}, X.~{Sun}, A.E.~{Bartels}, L.~{Ramos-Izquierdo},
  D.J.~{Krebs} et~al., \emph{{The Mercury Laser Altimeter Instrument for the
  MESSENGER Mission}},
  \href{https://doi.org/10.1007/s11214-007-9273-4}{\emph{\ssr} {\bfseries 131}
  (2007) 451}.

\bibitem{DSAC2021}
E.A.~Burt, J.D.~Prestage, R.L.~Tjoelker, D.G.~Enzer, D.~Kuang, D.W.~Murphy
  et~al., \emph{Demonstration of a trapped-ion atomic clock in space},
  \href{https://doi.org/10.1038/s41586-021-03571-7}{\emph{Nature} {\bfseries
  595} (2021) 43}.

\bibitem{Will2014}
C.M.~Will, \emph{The confrontation between general relativity and experiment},
  \href{https://doi.org/10.12942/lrr-2014-4}{\emph{Living Reviews in
  Relativity} {\bfseries 17} (2014) }.

\bibitem{Sch09}
S.~{Schiller}, G.M.~{Tino}, P.~{Gill}, C.~{Salomon}, U.~{Sterr}, E.~{Peik}
  et~al., \emph{{Einstein Gravity Explorer-a medium-class fundamental physics
  mission}},
  \href{https://doi.org/10.1007/s10686-008-9126-5}{\emph{Experimental
  Astronomy} {\bfseries 23} (2009) 573}.

\bibitem{Lit2021}
D.~{Litvinov} and S.~{Pilipenko}, \emph{{Testing the Einstein equivalence
  principle with two Earth-orbiting clocks}},
  \href{https://doi.org/10.1088/1361-6382/abf895}{\emph{Classical and Quantum
  Gravity} {\bfseries 38} (2021) 135010}
  [\href{https://arxiv.org/abs/2108.09723}{{\ttfamily 2108.09723}}].

\bibitem{Derevianko:2013oaa}
A.~Derevianko and M.~Pospelov, \emph{{Hunting for topological dark matter with
  atomic clocks}}, \href{https://doi.org/10.1038/nphys3137}{\emph{Nature Phys.}
  {\bfseries 10} (2014) 933} [\href{https://arxiv.org/abs/1311.1244}{{\ttfamily
  1311.1244}}].

\bibitem{Wcislo:2018ojh}
P.~Wcislo et~al., \emph{{New bounds on dark matter coupling from a global
  network of optical atomic clocks}},
  \href{https://doi.org/10.1126/sciadv.aau4869}{\emph{Sci. Adv.} {\bfseries 4}
  (2018) eaau4869} [\href{https://arxiv.org/abs/1806.04762}{{\ttfamily
  1806.04762}}].

\bibitem{Derevianko:2021wgw}
A.~Derevianko, D.J.~Kimball and C.~Dailey, \emph{{Reply to the comment on
  ''Quantum sensor networks as exotic field telescopes for multi-messenger
  astronomy''}},  \href{https://arxiv.org/abs/2112.02653}{{\ttfamily
  2112.02653}}.

\bibitem{Shen:2021scn}
W.~Shen et~al., \emph{{Testing gravitational redshift base on microwave
  frequency links onboard China Space Station}},
  \href{https://arxiv.org/abs/2112.02759}{{\ttfamily 2112.02759}}.

\bibitem{inproceedings}
A.~Batista, E.~Gomez, H.~Qiao and K.~Schubert, \emph{Constellation design of a
  lunar global positioning system using cubesats and chip-scale atomic clocks},
   07, 2012.

\bibitem{Adler:2008rq}
S.L.~Adler, \emph{{Placing direct limits on the mass of earth-bound dark
  matter}}, \href{https://doi.org/10.1088/1751-8113/41/41/412002}{\emph{J.
  Phys.} {\bfseries A41} (2008) 412002}
  [\href{https://arxiv.org/abs/0808.0899}{{\ttfamily 0808.0899}}].

\end{thebibliography}\endgroup

\appendix

\section{Properties of a Bound Solar Halo} \label{app:constraints}

\begin{figure}[t!]
 \includegraphics[scale=0.9]{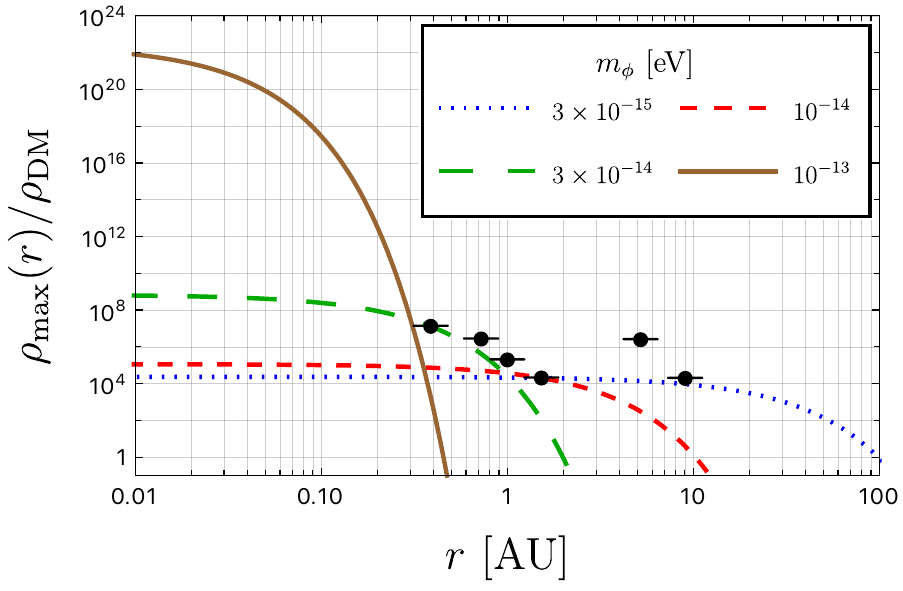}
 \caption{
 The maximum allowed density $\rho_{\rm max}(r)$ relative to background DM density $\rho_{\rm DM}$ as a function of distance from the Sun $r$. The blue dotted, red dashed, green long-dashed, and brown solid lines correspond to ULDM particle masses of $m_\phi = 3\times10^{-15}$ eV, $10^{-14}$ eV, $3\times10^{-14}$ eV, and $10^{-13}$ eV, respectively. The black points denote the constraints at the orbital radii of Mercury, Venus, Earth, Mars, Jupiter, and Saturn (left to right in the Figure) \cite{Pitjev:2013sfa}.
 }
 \label{fig:rhoofr}
\end{figure}

Under the usual assumptions, DM exists in a virialized configuration with roughly constant density $\rho_{DM}=0.4$ GeV/cm$^3$ in our solar neighborhood. The strongest local constraints arise from the orbital dynamics of planets in the solar system, i.e., solar system ephemerides; observations constrain the density of DM at the orbital radius of Mercury, Venus, Earth, Mars, Jupiter, and Saturn at the level of $\rho \lesssim 10^{3}-10^5$ GeV/cm$^3$ \cite{Pitjev:2013sfa}, which are shown by the black dots in Fig. \ref{fig:rhoofr}.

We have been considering the scenario in which ULDM fields become bound to objects in the solar system, in which case the density and coherence properties will be modified in ways that are relevant to experimental searches \cite{Banerjee:2019epw,Banerjee:2019xuy}. A bound solar halo is essentially similar to a gravitational atom, with the Sun playing the role of the nucleus; therefore the solar halo density function can be approximated as an exponential 
\begin{equation} \label{eq:rhostar}
 \rho(r) \simeq \rho_{\star}\exp\left(-2r/R_\star\right)    
\end{equation}
as long as $r\gg R_\star \gg R_\odot$, in precise analogy to the ground state of a hydrogen atom. As explained in the main text, the radius $R_\star$ is fully determined by its host mass $M_{\rm ext}$ (in the case under consideration, $M_{\rm ext} = M_\odot$) and the ULDM particle mass $m_\phi$; see Eq.~\eqref{eq:Rstar}. Therefore density function $\rho(r)$ for a SH is fully calculable, given input values of scalar mass $m_\phi$, radius $r$, and overall density normalization $\rho_\star$. For a density that saturates the limits of Ref. \cite{Pitjev:2013sfa}, we show the resulting density function $\rho(r)$, relative to $\rho_{\rm DM}$, for a few relevant choices of $m_\phi$, in Fig. \ref{fig:rhoofr}.

\begin{figure}[t!]
 \includegraphics[scale=0.9]{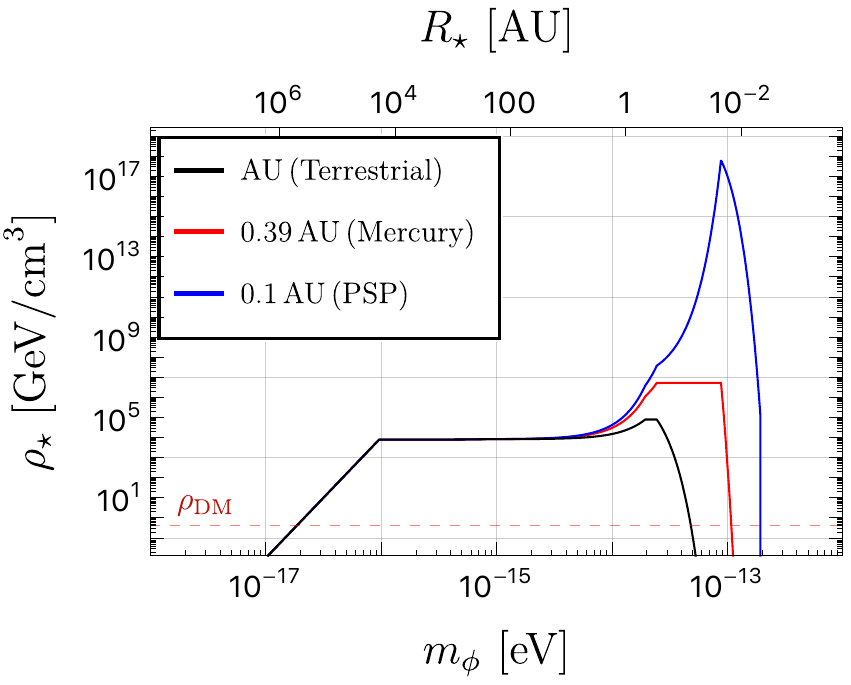}\quad
 \includegraphics[scale=0.9]{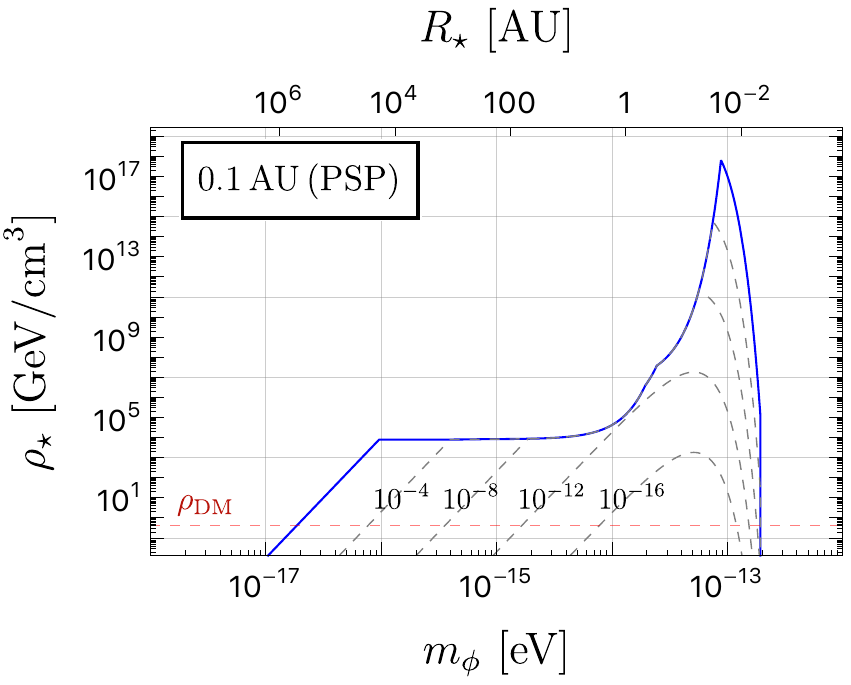}
 \caption{
 The allowed solar halo density $\rho_\star$ as a function of ULDM particle mass $m_\phi$.
 \underline{Upper panel:} Maximum density $\rho_\star$ at different probe radii: AU (radius of Earth's orbit, black line), $0.39$ AU (radius of Mercury's orbit, red line), and $0.1$ AU (blue line). Burgundy dashed line is the local density of virialized DM.
 \underline{Lower panel:} Maximum density at $0.1$ AU (blue line), along with contours of fixed $M_\star/M_\odot$ as labeled.
 }
 \label{fig:density}
\end{figure}

\begin{figure}[t]
 \includegraphics[scale=0.58]{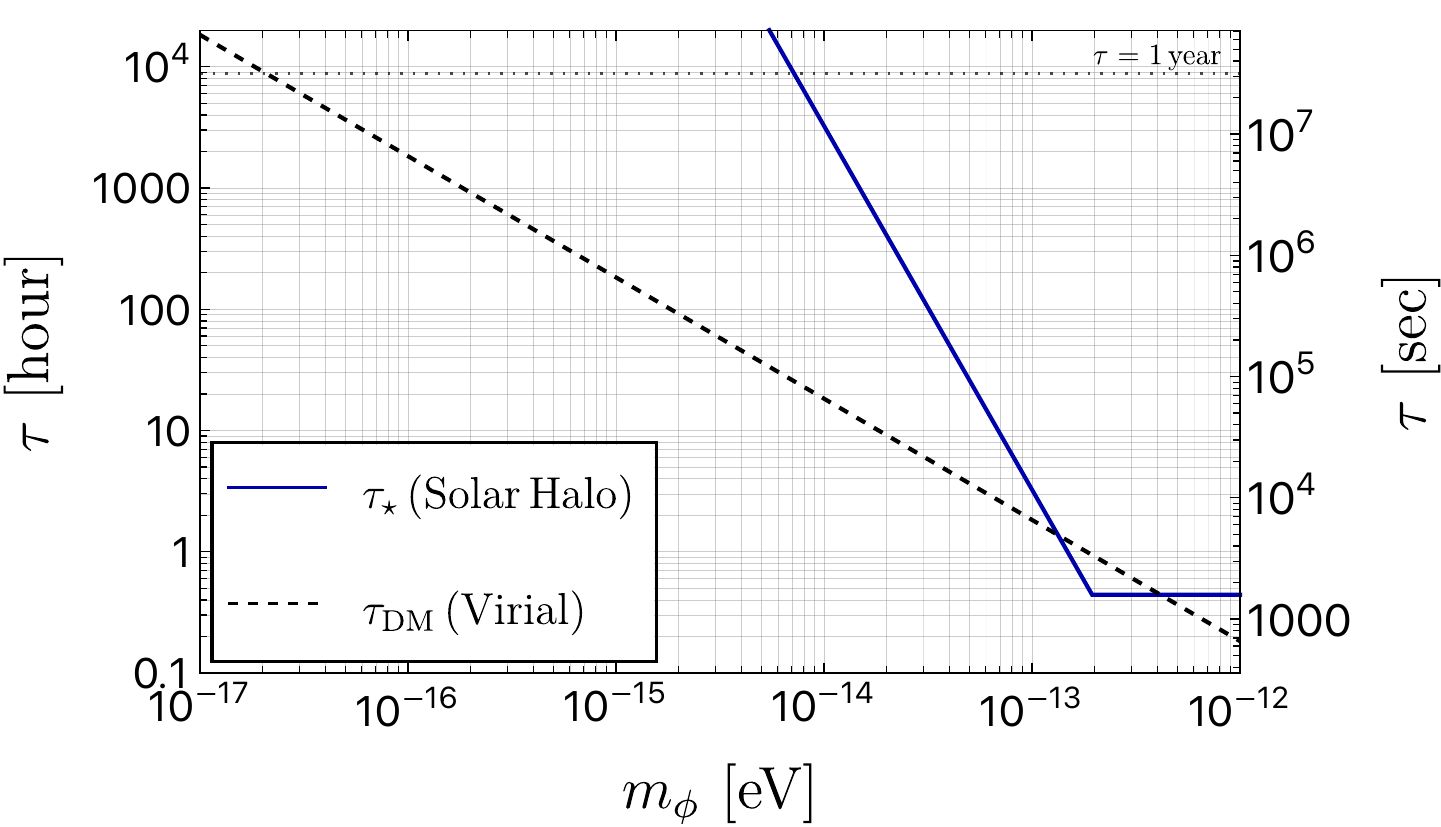}
 \caption{
 The coherence timescale for a solar halo (blue) and for virialized DM (black dashed), as a function of ULDM particle mass $m_\phi$.
 }
 \label{fig:tcoh}
\end{figure}

Given the above discussion of solar halo properties, we can translate the constraints on the local DM density from Ref. \cite{Pitjev:2013sfa} into a constraint on $\rho_{\star}$ as a function of $m_\phi$. This is what is shown in the upper panel of Fig. \ref{fig:density}: we have illustrated the resulting limits on solar halo density $\rho_\star$ at a distance $1$ AU from the sun (relevant for Earth-based probes, black line), a distance of $0.39$ AU (average radius of Mercury's orbit, red line), and for a distance of $0.1$ AU (blue line). In the lower panel, we illustrate the same maximum density at $0.1$ AU, but including contours to show the density for different choices of $M_\star/M_\odot$. Note that for consistency, we enforce $M_\star<M_\odot/2$ over the full range of parameters.
It is evident from the Figure that the constraint on the local density of DM bound to the Sun becomes very weak when measured inside the orbit of Mercury. In the lower panel, we observe that even a very small bound mass, of order $10^{-16} M_\odot$, can give rise to a $10^4$ increase in the density of DM at $0.1$ AU for $m_\phi \simeq 10^{-13}$ eV.

To estimate the experimental reach of an atomic clock to probe a SH, we saturate the constraint on SH density at a given radius $r$ (the upper panel of Fig. \ref{fig:density}). This translates into a field amplitude $\phi_\star = \sqrt{2\rho_\star}/m_\phi$, which we substitute in Eq. \eqref{eq:mualpha}. Then, we fix a value for the fractional accuracy in the variation of the fundamental constants
(e.g. $10^{-14}$) motivated by current and near-future experiments (see Section \ref{sec:clocks}), and derive the resulting sensitivity reach using
\begin{align}
 d_{m_e}^{\rm limit} &\simeq \frac{1}{\k\phi_\star(r)}\left(\frac{\d\m}{\m}\right)_{\rm exp}, \\
  d_{\a}^{\rm limit} &\simeq \frac{1}{\k\phi_\star(r)}\left(\frac{\d\a}{\a}\right)_{\rm exp}, \\
  d_{g}^{\rm limit} &\simeq
  \frac{1}{\k\phi_\star(r)}
            \left(\frac{\d(m_q/\L_{\rm QCD})}{(m_q/\L_{\rm QCD})}\right)_{\rm exp}.
\end{align}
The results are indicated by the black, red, and blue lines in Fig. \ref{fig:sens}. 

The coherence timescale for ULDM oscillations is typically much longer in a solar halo than it would be for virial DM $\tau_{\rm DM}$. This is because the bound ULDM particles much be colder, i.e. have lower velocity dispersion, to remain bound to the Sun. The velocity dispersion $v_\star$, and therefore the coherence timescale $\tau_\star$, of a SH is essentially dictated by its radius $R_\star$ and particle mass $m_\phi$ via the relation \cite{Banerjee:2019xuy}
\begin{align}
 \tau_\star &\simeq (m_\phi v_\star^2)^{-1} \simeq m_\phi R_\star^2 \nn \\
 &\simeq 10^3\,{\rm sec}\times 
 \begin{cases}
  1, & m_\phi \gtrsim 2\times10^{-13}\,{\rm eV} \\
  \left(\frac{\displaystyle 2\times10^{-13}\,{\rm eV}}{\displaystyle m_\phi}\right)^3, & m_\phi \lesssim 2\times10^{-13}\,{\rm eV}.
 \end{cases}
\end{align}
This relation is illustrated in Fig. \ref{fig:tcoh}.
When $\tau_\star$ is much longer than the averaging period $\tau$ of an atomic clock DM search (see discussion around Eq. \eqref{sigma}), the full stability of the clock can be leveraged; on the other hand, when $\tau_\star < \tau$, the sensitivity to ULDM signals is diminished by the factor $\sqrt{\tau_\star/\tau}$. For searches that are shorter than $\tau\simeq 1$ day, the resulting reduction in sensitivity is about one order of magnitude at worst, when $m_\phi \gtrsim 2\times10^{-13}$ eV.

\section{Clocks in Earth Orbits and An Earth-bound Halo} \label{app:earthclock}

It has been proposed to put atomic clocks in orbit around the Earth \cite{Sch09,Lit2021} and on the moon (see e.g. \cite{inproceedings}). For completeness, we note here that depending on the details of the clock, such probes could be sensitive to ULDM bound halos around the Earth. Because of the dependence of the halo radius on $m_\phi$ (see Eq.~\eqref{eq:Rstar}), such probes would be restricted to a mass window of larger $m_\phi$, with peak sensitivity around $10^{-11}$ eV and $10^{-8}$ eV.
Atomic spectroscopy searches in this range are rapidly progressing \cite{Aharony:2019iad,Antypas:2019qji,Savalle:2020vgz}, and molecular spectroscopy experiments are able to achieve sensitivity at the level of $10^{-15}$ \cite{Oswald:2021vtc}.
We also note that in a scenario with a large quadratic coupling to matter and resulting screening of the ULDM field in the vicinity of the Earth \cite{Hees:2018fpg}, a probe in orbit may be advantageous in evading this screening. It is worthwhile to consider the possible sensitivity of these probes as well.

For a halo bound to the Earth, the situation is similar to that of the SH in Appendix \ref{app:constraints}, with the substitution $M_{\rm ext} = M_\oplus$ and the relevant mass range is at larger $m_\phi \gtrsim 10^{-13}$ eV based on Eq.~\eqref{eq:Rstar}. The strongest constraints on this scenario arise from comparison of orbital dynamics of low-orbit satellites (e.g. LAGEOS) and the moon \cite{Adler:2008rq}. See \cite{Banerjee:2019epw} for a more complete treatment of this case.

\begin{figure*}[t]
\begin{minipage}[c]{0.45\textwidth}
\hspace{-7 cm} \large{(a)} \\
\vspace{-0.5cm}
 \includegraphics[scale=0.9]{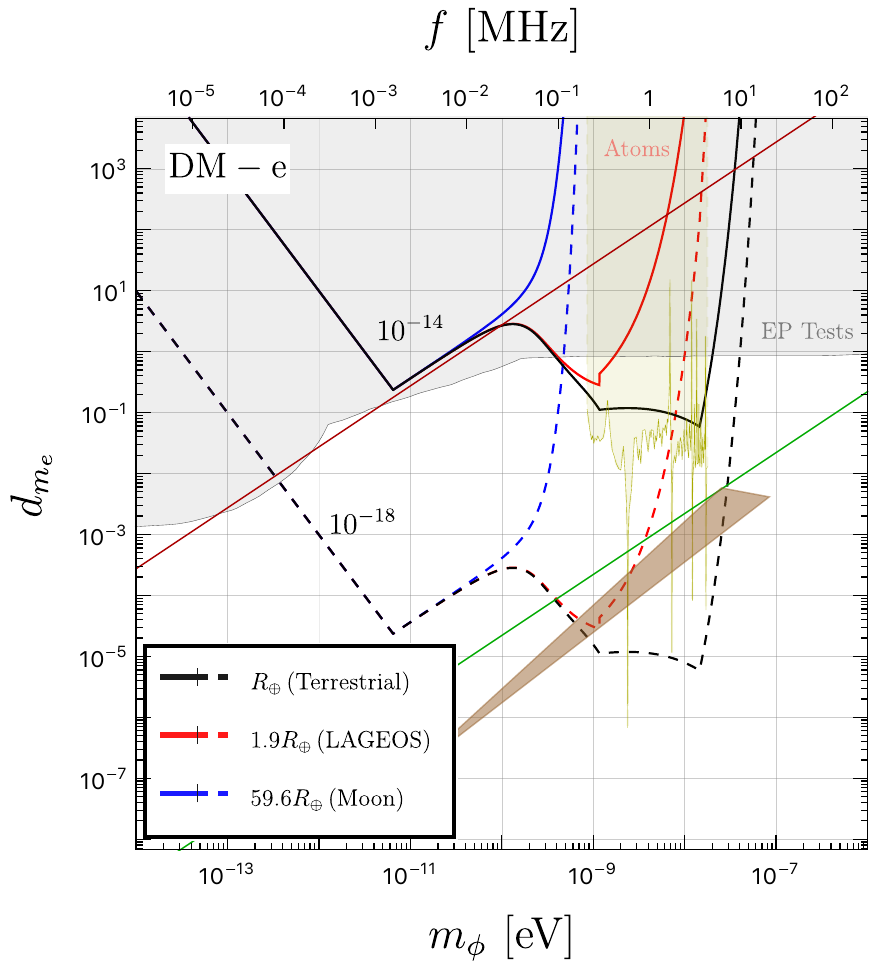} \quad
\end{minipage}
\begin{minipage}[c]{0.45\textwidth}
\vspace{-0.4cm}
 \hspace{-7 cm} \large{(b)} \\
\vspace{-0.5cm}
 \includegraphics[scale=0.9]{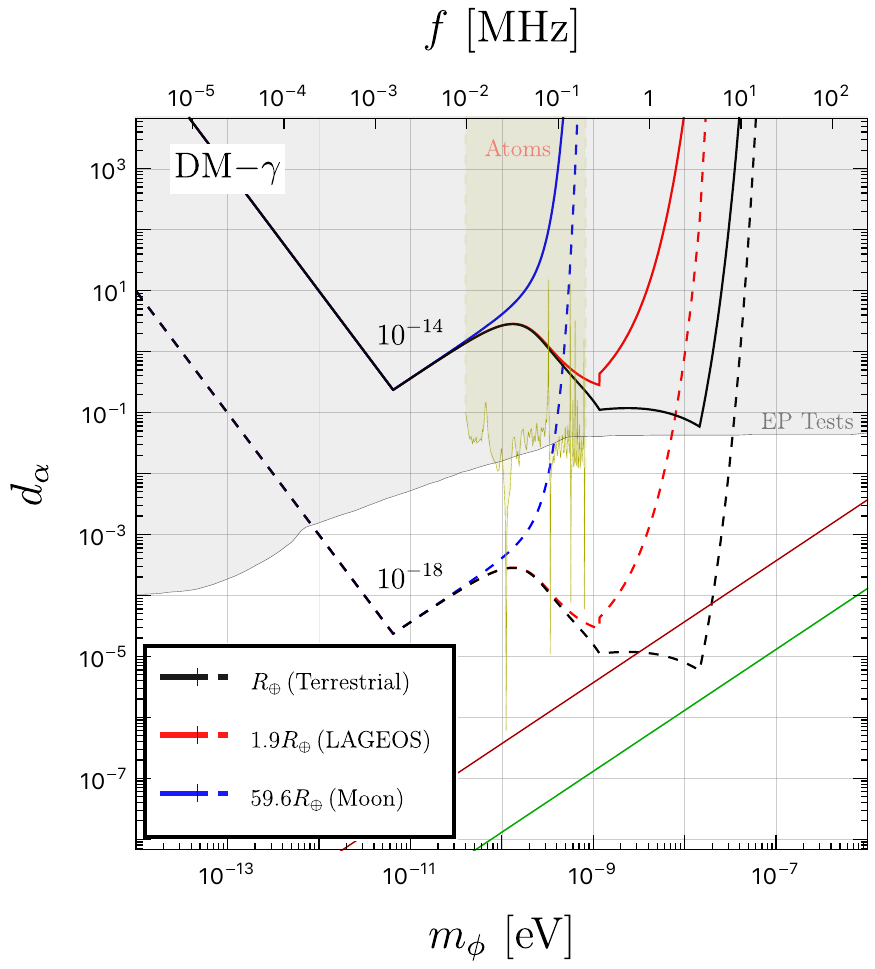}
\end{minipage}
 \caption{Estimated sensitivity reach for ultralight dark matter, coupled via Eq. \eqref{eq:Lag} and bound to the Earth; for the bound Earth halo density, we assume the maximum allowed by gravitational constraints following \cite{Banerjee:2019epw}. The blue, red, and black denote sensitivity for those at distances explored by the Moon, the LAGEOS Satellite, and terrestrial clocks, respectively. The thick and dashed lines correspond to assumed sensitivity at the level of $10^{-14}$ and $10^{-18}$ (respectively) to variations of (a) $(\delta \mu/\mu)$ or (b) fine structure constant $(\d\a/\a)$.
The gray and yellow regions denote the current constraint from equivalence principle tests \cite{Wagner:2012ui,Berge:2017ovy} and atomic physics probes of an Earth Halo \cite{Savalle:2020vgz} (respectively); the diagonal burgundy and green lines denote motivated theory targets, as in Fig. \ref{fig:sens} \cite{Flacke:2016szy,Choi:2016luu}. The brown shaded region is a target for probing coherent relaxion dark matter \cite{Banerjee:2018xmn}.
}
 \label{fig:sensMoon}
\end{figure*}

In Fig. \ref{fig:sensMoon} we illustrate the sensitivity of an atomic clock to the presence of a ULDM halo around the Earth. The style of the lines is identical to that of Fig. \ref{fig:sens}, except that the red and blue lines correspond to probes at a distance equal to the orbit of the LAGEOS Satellite, $1.9R_\oplus$ \cite{Adler:2008rq}, and that of the moon, $59.6R_\oplus$. We have assumed a sensitivity of $10^{-14}$ (thick lines) or $10^{-18}$ (dashed) to the oscillating signal, as in Fig. \ref{fig:sens}. 
We found that a terrestrial clock is sufficient in probing the Earth-bound halo. Placing clocks in LAGEOS orbit or on the Moon results in reduced sensitivities for the Earth-bound halo, assuming there are no ULDM-SM quadratic couplings and the associated Earth-screening effects.

As we have noted throughout the paper, the sensitivity to the bound halo scenario is \emph{diminished} when the probe is sent far from the center of the halo, due to the halo profile and the constraints on the halo density. In the main text, we discussed the possibility of a probe being sent towards the Sun, which would allow one to probe weaker couplings, and larger masses $m_\phi$, of particles in the bound SH. An analogous case for an EH would suggest a mission to send an atomic clock being down into the Earth as deeply as possible. The largest depth ever probed by humankind is only of order $10$ km below the surface, where the increase in sensitivity would be invisible at the scale of Fig. \ref{fig:sensMoon}. One would require a clock with $10^{-18}$ accuracy poised a distance $2000$ km below the surface (in the mantle of the Earth) to probe bound scalar ULDM up to masses as high as $10^{-7}$ eV, which is clearly not viable.

\end{document}